\documentclass[12pt]{article}
\usepackage{cancel}
\usepackage{rotating}
\usepackage{booktabs}
\usepackage{mathrsfs}
\usepackage{fancyhdr}
\usepackage{amsfonts}
\usepackage{amstext}
\usepackage{amsmath}
\usepackage{amssymb}
\usepackage{extarrows}
\usepackage{color,xcolor}
\usepackage[a4paper,includeheadfoot,nomarginpar]{geometry}
\usepackage{setspace}
\usepackage{longtable}
\usepackage{float}
\usepackage{subfigure}
\usepackage{fullpage}
\usepackage{times}
\usepackage[utf8]{inputenc}
\usepackage{units}
\usepackage{graphics}
\usepackage{nicefrac}
\usepackage{color}
\usepackage{amsthm}
\RequirePackage{natbib}
\RequirePackage[colorlinks,citecolor=blue,urlcolor=blue]{hyperref}

\newtheorem{innercustomgeneric}{\customgenericname}
\providecommand{\customgenericname}{}
\newcommand{\newcustomtheorem}[2]{%
  \newenvironment{#1}[1]
  {%
   \renewcommand\customgenericname{#2}%
   \renewcommand\theinnercustomgeneric{##1}%
   \innercustomgeneric
  }
  {\endinnercustomgeneric}
}

\newcommand{\R}{\mathbb{R}}
\newcommand{\B}{\mathbb{B}}
\newcommand{\pr}{\mathrm{P}}
\newcommand{\E}{\mathrm{E}}
\newcommand{\F}{\mathcal{F}}
\newcommand{\N}{\mathcal{N}}
\newcommand{\W}{\mathcal{W}}

\newcommand{\Y}{\mathcal{Y}}
\newcommand{\Z}{\mathcal{Z}}

\newcustomtheorem{customlemma}{Lemma}
\newtheorem{theorem}{Theorem}

\newtheorem{corollary}{Corollary}

\newtheorem{remark}{Remark}

\begin{document}

\title{Nonparametric Tests of Conditional Independence for Time Series}

\author{Xiaojun Song \ \thanks{Department of Business Statistics and Econometrics, Guanghua School of
Management and Center for Statistical Science, Peking University, Beijing, 100871, China. E-mail: sxj@gsm.pku.edu.cn. Financial support from the National Natural Science Foundation of China (Grant No. 71532001) is acknowledged.}\\Peking University
\and
Haoyu Wei \thanks{Guanghua School of Management, Peking University, Beijing, China. Email: cute@pku.edu.cn. }\\Peking University
}

\renewcommand{\thefootnote}{\fnsymbol{footnote}}
%\footnotetext[3]{The two authors contributed equally to this work in alphabetic order.}
% \footnotemark[3] use this to add equal work
\maketitle

\begin{abstract}
We propose consistent nonparametric tests of conditional independence for time series data. Our methods are motivated from the difference between joint conditional cumulative distribution function (CDF) and the product of conditional CDFs. The difference is transformed into a proper conditional moment restriction (CMR), which forms the basis for our testing procedure. Our test statistics are then constructed using the integrated moment restrictions that are equivalent to the CMR. We establish the asymptotic behavior of the test statistics under the null, the alternative, and the sequence of local alternatives converging to conditional independence at the parametric rate. Our tests are implemented with the assistance of a multiplier bootstrap. Monte Carlo simulations are conducted to evaluate the finite sample performance of the proposed tests. We apply our tests to examine the predictability of equity risk premium using variance risk premium for different horizons and find that there exist various degrees of nonlinear predictability at mid-run and long-run horizons.
\end{abstract}



\thispagestyle{empty}

\newpage \pagenumbering{arabic} \setcounter{page}{1} \setcounter{footnote}{0}
%TCIMACRO{%
%\TeXButton{\arabic{footnote}}{\renewcommand{\thefootnote}{\arabic{footnote}}}}%
%BeginExpansion
\renewcommand{\thefootnote}{\arabic{footnote}}%
%EndExpansion
%\begin{doublespace}

\pagebreak

\section{Introduction}
A variable $Y$ is said to be conditionally independent of $Z$ given $X$ if and only if the conditional density of $Y$ given $Z$ and $X$ equals to the conditional density of $Y$ given $X$, that is, $Z$ does not carry any information about $Y$ once $X$ is given. Following David (1979), we write $Y\bot Z|X$ to denote that $Y$ is independent of $Z$ given $X$. The hypotheses $Y\bot Z|X$ is related to the hypothesis that $Y$ is independent of $Z$, i.e. $Y\bot Z$ (unconditional independence), and the conditional mean independence (CMI), $\E (Y|Z,X) = \E (Y|X)$. 

The assumption of conditional independence plays an important role and is a widely imposed one in both statistical and econometric literature. For example, Markov property of a time series process, Granger non-causality, the assumption of missing at random (MAR) and exogeneity all can be formulated as a conditional independence restriction, see \cite{hong2017testing} for motivating examples of testing the conditional independence hypothesis in economics and econometrics. However, in contrast to the many tests of unconditional independence or CMI proposed in the context of independent and identically distributed (i.i.d.) data, by far, not many tests are available for testing conditional independence assumption with time series data. Nonparametric tests for unconditional independence between random variables and/or vectors, and nonparametric tests for serial independence are abundant, e.g. a nonparametric test of Cram\'{e}r-von Mises type first introduced by \cite{hoeffding1948non}, the empirical distribution function-based tests of \cite{blum1961distribution}, \cite{skaug1993nonparametric} for testing independence of raw data and \cite{delgado2000nonparametric} or \cite{ghoudi2001nonparametric} for testing serial independence of time series or regression errors,  the empirical characteristic function-based test of \cite{csorgHo1985testing}, kernel smoothing-based tests like \cite{rosenblatt1975quadratic}, \cite{robinson1991blup}, and \cite{hong2005asymptotic}, and tests based on measures of association and dependence between random variables and/or vectors such as \cite{bakirov2006multivariate}, \cite{szekely2007measuring} or \cite{diks2007nonparametric}. On the other hand, among the available tests for conditional independence, the majority is designed for i.i.d. data, e.g. \cite{linton2014testing} develop a non-pivotal nonparametric test based on a generalization of the empirical distribution function, \cite{song2009testing} employs the Rosenblatt transformation to obtain a distribution-free test for a different type of conditional independence, \cite{huang2010testing} proposes a test based on the maximal nonlinear conditional correlation, and \cite{huang2016flexible} develop an integrated conditional moment test. Tests suitable for time series data include \cite{su2007consistent, su2008nonparametric, su2012conditional, su2014testing}, \cite{bouezmarni2012nonparametric} and \cite{wang2018characteristic}, all of which are based on kernel smoothing. The exception is \cite{su2012conditional}, who provide nonparametric tests for conditional independence using local polynomial quantile regression. 

In this paper we aim to further fill the gap of the literature and propose consistent nonparametric tests based on an empirical process approach for testing conditional independence which are applicable to time series data. Our approach exploits a proper conditional moment restriction and is in the same spirit with \cite{delgado2001significance}, which partially circumvents the ``curse of dimensionality'' problem. Besides, in comparison with the existing tests based on smoothing methods, our new tests are able to detect local alternatives converging to the null at a parametric rate. 

We introduce some notations for testing the conditional independence hypothesis in a time series framework. Let $X_t$, $Y_t$ and $Z_t$ be three generic random vectors with dimensions $d_x$, $d_y$ and $d_z$, respectively. The null hypothesis of interest is that $Y_t$ is independent of $Z_t$ conditional on $X_t$, i.e.,
\begin{equation*}
F_{Y,Z|X}(y,z|x)=F_{Y|X}(y|x)F_{Z|X}(z|x),
\end{equation*}
for all $(x,y,z)\in\mathbb{R}^{d_x+d_y+d_z}$, where $F_{Y,Z|X}$, $F_{Y|X}$ and $F_{Z|X}$ denote the conditional cumulative distribution functions (CDFs).

The rest of the paper is as follows. In Section 2 we examine the testing problem and provide the test statistics. Section 3 establishes the asymptotic null distributions. Section 4 studies the consistency property of the test and the asymptotic local power of the test under local alternatives. A bootstrap procedure to implement the tests is proposed and formally justified in Section 5. In Section 6, we study the finite sample performance of our tests by means of Monte Carlo simulations. Section 7 presents an empirical example of using variance risk premium to predict equity risk premium. Finally, Section 8 concludes the paper. All proofs are collected in the online Appendix.

\section{The testing procedure}\label{section2}
Consider a $\mathbb{R}^{d_x+d_y+d_z}$-valued strictly stationary ergodic time series process $\{(X_t,Y_{t},Z_{t})\}$ defined on the probability space $(\Omega ,\mathcal{F},\mathbb{P})$, which satisfies the Markov's property 
\begin{equation}
    \pr (Y_{t}\leq y,Z_t\leq z\vert\mathcal{A}_{t-1}) = \pr (Y_{t}\leq y,Z_t\leq z\vert W_{t})\,\,\,\text{a.s.}\,\,\,\forall(y,z)\in \mathbb{R}^{d_y+d_z}, \label{markov}
\end{equation}
where $\mathcal{A}_{t-1}:=\sigma(\{X_{s},Y_{s-1},Z_{s-1}\}_{s=-\infty}^{t})$ with $\sigma(\cdot)$ the smallest sigma algebra,
\begin{equation*}
    W_{t}=\{(X^\top_s,Y^\top_{s-1},Z^\top_{s-1})^\top\}_{s=t-p+1}^t
\end{equation*}
with $0<p<\infty$ an integer, and ``$^\top$'' denotes transpose. That is, the only relevant information for explaining $(Y_{t},Z_t)$ are the first $m$ lags of $\left(X_{t+1}, Y_{t},Z_{t}\right)$. Note that $W_t$ is a $d_w\times 1$ vector with $d_w=p(d_x+d_y+d_z)$.

We propose a nonparametric test for the hypothesis that $Y_t$ and $Z_t$ are independent given $W_t$, i.e.,
\begin{equation}
    \text{H}_0: F_{Y,Z|W}(y,z|W_t)=F_{Y|W}(y|W_t)F_{Z|W}(z|W_t)\quad\text{a.s.} \quad \forall \,  (y,z)\in\mathbb{R}^{d_y+d_z}.\label{eq:null1}
\end{equation}
The alternative hypothesis $\text{H}_1$ is the negation of $\text{H}_0$ in \eqref{eq:null1}. The alternative hypothesis $\text{H}_1$ consists of a broad class of conditional dependence between $Y_t$ and $Z_t$ given $W_t$. That is, conditional on $W_t$, $Y_t$ could depend on $Z_t$ through mean, variance, skewness, kurtosis, or even higher moments. It is possible to have situations where the dependence between $Y_t$ and $Z_t$ in lower moments (e.g. mean or variance) does not exist, but it does exist in higher moments (e.g. skewness or kurtosis). See Section \ref{mc} for some data generating processes under $\text{H}_1$.

It is important to emphasize that the new formulation of testing conditional independence assumption in \eqref{eq:null1} is attractive, since it partly circumvents the problem of ``curse of dimensionality'' by conditioning on only $W_t$ in all three conditional CDFs. \cite{wang2018characteristic} has also exploited \eqref{eq:null1} to propose a test based on the conditional characteristic functions. On the other hand, existing tests are mainly based on testing $F_{Y|W,Z}(y|W_t,Z_t)=F_{Y|W}(y|W_t)$ a.s., which requires estimation of conditional CDFs given both $W_t$ and $Z_t$, see e.g. \cite{su2007consistent, su2008nonparametric} and \cite{bouezmarni2012nonparametric} to name only a few.  

Note that $F_{Y,Z|W}(y,z|W_t) = \E[1(Y_t\leq y)1(Z_t\leq z)|W_t]$ and $F_{Y|W}(y|W_t) = \E[1(Y_t\leq y)|W_t]$, with $1(\cdot)$ the indicator function. Thus, $\text{H}_0$ in \eqref{eq:null1} can be tested by
\begin{equation}
    \text{H}_0: \E [1(Y_t\leq y)(1(Z_t\leq z)-F_{Z|W}(z|W_t))|W_t]=0 \quad \text{a.s.} \quad \forall \, (y,z)\in\mathbb{R}^{d_y+d_z}. \label{eq:null3}
\end{equation}
The problem of testing \eqref{eq:null1} is in fact equivalent to testing a conditional moment restriction (CMR) in \eqref{eq:null3}, which has the advantage of conditioning on only $W_t$. An important feature of the CMR stated in \eqref{eq:null3} is that it involves all values of $(y,z)$. Since it has to hold for all $(y,z)$, we have an infinite number of CMRs to be tested.

Fortunately, \cite{stinchcombe1998consistent} give us a method to convert conditional to unconditional moment restriction in a convenient way: testing \eqref{eq:null3} is further equivalent to testing the following infinite number of unconditional moment restrictions,
\begin{equation}\label{eq:null4}
    \text{H}_0: \E \big[ \varphi(W_t, w) 1 (Y_t \leq y) \big( 1 (Z_t \leq z) - F_{Z | W} (z | W_t)\big) \big] = 0, \quad \forall \, (w, y, z) \in \W \times \mathbb{R}^{d_y+d_z}, 
\end{equation}
where $\W \subseteq \mathbb{R}^{d_{w}}$ is a proper chosen set with typical choices $d_{w} = d_x + d_y + d_z$ or $d_{w} = d_x + d_y + d_z + 1$, and $\varphi$
is a generically comprehensively revealing (GCR) or comprehensively revealing (CR) function. Examples of GCR functions include: (1) $\varphi(W_t, w) = \exp (\mathrm{i} w^{\top} W_t)$; (2) $\varphi(W_t, \gamma) = \sin (w^{\top} W_t)$, and examples of CR functions include: (3) $\varphi(W_t, \gamma) = 1 (W_t \leq w)$; (4) $\varphi (W_t, w) = 1 (\beta^{\top} W_t \leq \alpha)$ with $w = (\alpha, \beta)^{\top}$. It is worthy to note that when $\varphi$ is GCR, the deviations from the null hypothesis can be detected by essentially any choice of $w \in \W$, where $\W$ can be chosen as any small compact set with non-empty interior, whereas CR functions may require the set $\W$ to be the whole Euclidean space to ensure the consistency of the associate test. More discussion can be seen in \cite{stinchcombe1998consistent} and \cite{su2012conditional}. Hence in the following, we will always assume $\W$ is a bounded space in $\mathbb{R}^{d_w}$. In addition, to avoid the random denominator problem, we propose to test the following modified version of \eqref{eq:null4},
\begin{equation}\label{eq:null5}
    \text{H}_0: \E \big[ \varphi(W_t, w) 1 (Y_t \leq y) \big( 1 (Z_t \leq z) - F_{Z | W} (z | W_t)\big) f_W (W_t)\big] = 0, \quad \forall \, (w, y, z) \in \W \times \mathbb{R}^{d_y+d_z}, 
\end{equation}
where $f_W(W_t)$ is the density of $W_t$. As in \cite{delgado2001significance}, the density-weighted formulation helps to avoid conveniently the random denominator in the subsequent nonparametric estimation.

Given a sample $\{(W^\top_t,Y^\top_t,Z^\top_t)^\top\}_{t=1}^n$ of size $n$, if $F_{Z|W}(z|W_t)$ and $f_W(W_t)$ were observable, test statistics could be constructed using the (infeasible) empirical process
\begin{align*}
    S^{0}_{n}(w, y, z)=\frac{1}{\sqrt{n}}\sum_{t=1}^n \varphi(W_t, w) 1(Y_t\leq y)(1(Z_t\leq z)-F_{Z|W}(z|W_t))f_W(W_t).
\end{align*}
Then under $\text{H}_0$, $S^{0}_{n}(\cdot,\cdot,\cdot) \rightsquigarrow S^{0}_\infty(\cdot,\cdot,\cdot)$, where $S^{0}_\infty(\cdot,\cdot,\cdot)$ is a zero mean Gaussian process with covariance kernel $\E\left[S^{0}_\infty(w,y,z)S^{0}_\infty(w',y',z')\right]$.

As $F_{Z|W}(z|W_t)$ and $f_W(W_t)$ are unobservable, testing procedures based on $S^0_{n}(w,y,z)$ are not feasible. In this paper, we replace them with $\widehat{F}_{Z|W}(z|W_t)$ and $\widehat{f}_W(W_t)$, where 
\begin{equation*}
    \widehat{F}_{Z|W}(z|W_t)=\frac{\frac{1}{(n-1)h^{d_w}}\sum_{s=1, s\neq t}^n1(Z_s\leq z)K\left(\frac{W_t-W_s}{h}\right)}{\hat{f}_W(W_t)},
\end{equation*}
\begin{equation*}
    \widehat{f}_W(W_t)=\frac{1}{(n-1)h^{d_w}}\sum_{s=1, s\neq t}^nK\left(\frac{W_t-W_s}{h}\right),
\end{equation*}
with $K(\cdot)$ and $h:=h_n\in\mathbb{R}^+$ the kernel function and bandwidth, respectively. Define the feasible empirical process
\begin{align*}
    S_{n}(w,y,z)=\frac{1}{\sqrt n}\sum_{t=1}^n \varphi(W_t, w) 1(Y_t\leq y)(1(Z_t\leq z)-\widehat F_{Z|W}(z|W_t))\widehat f_W(W_t),
\end{align*} 
which is algebraically equivalent to 
\begin{equation*}
    \frac{1}{\sqrt n(n-1)h^{d_w}}\sum_{t=1}^n\sum_{s=1, s\neq t}^nK\left(\frac{W_{t}-W_{s}}{h}\right) \varphi (W_t, w) 1(Y_t\leq y)(1(Z_t\leq z)-1(Z_s\leq z)).
\end{equation*} 
The above expression is a variant of $U$-processes considered by \cite{delgado2001significance} in an i.i.d. context. Note that the limiting distribution of $S_{n}(w,y,z)$ will be different from that of $S^0_{n}(w,y,z)$ due to the estimation of $F_{Z|W}(z|W_t)$.

Test statistics are constructed based on suitable continuous functionals of $S_n(w,y,z)$. A test statistic in the spirit of the Cram\'{e}r-von Mises type is
\begin{align}
    CvM_n=\int S^2_n(w,y,z)\,dF_{n}(w,y,z)=n^{-1}\sum_{t=1}^{n}S_n^2(W_t,Y_t,Z_t), \label{cvmn}
\end{align}
where $F_n(w,y,z)=n^{-1}\sum_{t=1}^n1(W_t\leq x)1(Y_t\leq y)1(Z_t\leq z)$ is the empirical distribution function of $(X_t,Y_t,Z_t)$.  Henceforth, an unspecified integral denotes integration over the whole space. The Kolmogorov-Smirnov type test statistic basing on the sup-norm is
\begin{equation}
KS_n=\sup_{(w,y,z) \in \W \times \mathbb{R}^{d_y} \times \mathbb{R}^{d_z}}\left\vert S_{n}(w,y,z)\right\vert. \label{ksn}
\end{equation}
In practice, one can compute $KS_n$ by simply taking the maximum over the observations, i.e., $\widetilde{KS}_{n}=\max_{1\leq t\leq n}\vert S_n(W_t,Y_t,Z_t)\vert$.

Under $\text{H}_0$, test statistics $CvM_n$ and $KS_n$ converge in distribution, while they diverge to infinity under $\text{H}_1$. We reject the null hypothesis of conditional independence whenever they exceed certain ``large'' values. Since the asymptotic null distributions of $CvM_n$ and $KS_n$ depend on the data generating process in a complicated way, their critical values are not readily available. To circumvent this problem, we propose a bootstrap procedure to obtain the critical values of our tests in Section \ref{boot}.

\section{Asymptotic null distributions}\label{section3}
In this section, we will establish the asymptotic null distributions of our test statistics $CvM_n$ and $KS_n$. We need to impose the following assumptions, which are attached in Appendix \ref{assumptions}. 

Let $\phi_t(y)=1(Y_{t}\leq y)-F_{Y|W}(y|W_t)$, $\epsilon_t(z)=1(Z_{t}\leq z)-F_{Z|W}(z|W_t)$, and $e_t(w,y,z) = \varphi (W_t, w) \phi_t(y)\epsilon_t(z)f_W(W_t)$. The following theorem shows that $S_n(\cdot,\cdot,\cdot)$ converges weakly to a Gaussian process under the null.

\begin{theorem}\label{thm1}
    Suppose \eqref{markov} and Assumption A1 - A7 in Appendix \ref{assumptions} hold. Then under the null
    \begin{equation*}
        S_n (\cdot, \cdot, \cdot) \rightsquigarrow S_{\infty} (\cdot, \cdot, \cdot),
    \end{equation*}
    where $S_{\infty} (\cdot, \cdot, \cdot)$ is a zero mean Gaussian process with covariance kernel $\E \big[ S_{\infty} (w, y, z) S_{\infty} (w', y', z') \big]$. Specially, if $(X_t^{\top}, Y_t^{\top} , Z_t^{\top})$ is IID sample, we have
    \begin{align*}
        \E\big[S_{\infty}(w,y,z), S_{\infty}(w',y',z')\big]& = \E \big[e_1(w,y,z)e_1(w',y',z')\big]\\
        & = \E[\varphi(W_t, w) \varphi(W_t, w') \psi(y,y';W_1)\rho(z,z';W_1)f^2_W(W_1)],
    \end{align*}
    where $\psi(y,y';W)=F_{Y|W}(y\wedge y'|W)-F_{Y|W}(y|W)F_{Y|W}(y'|W)$ and $\rho(z,z';W)=F_{Z|W}(z\wedge z'|W)-F_{Z|W}(z|W)F_{Z|W}(z'|W)$.
\end{theorem}

\begin{remark}
    The term $F_{Y|W}(y|W_t)$ in the definition of $e_t(w,y,z)$ reflects the cost paid for replacing $F_{Z|W}(z|W_t)$ with $\widehat F_{Z|W}(z|W_t)$ in the infeasible process $S_n^0(w,y,z)$, a phenomenon known as the ``parameter estimation error'', leading to a covariance kernel different from that of $S_n^0(w,y,z)$.
\end{remark}

The asymptotic null distributions of test statistics $CvM_n$ and $KS_n$  are given in the corollary below.

\begin{corollary}\label{cor1}
    Suppose \eqref{markov} and Assumption A1 - A7 in Appendix \ref{assumptions} hold. Then under the null,
    \begin{align*}
        CvM_n \rightsquigarrow CvM_{\infty}:=\int_{\W \times \mathbb{R}^{d_y} \times \mathbb{R}^{d_z}} S^2_{\infty}(w,y,z)\,dF_{W,Y,Z}(w,y,z),
    \end{align*}
    \begin{equation*}
        KS_n \rightsquigarrow KS_{\infty}:=\sup_{(w,y,z) \in \W \times \mathbb{R}^{d_y} \times \mathbb{R}^{d_z}}\left\vert S_{\infty}(w,y,z)\right\vert,
    \end{equation*}
    where $S_{\infty}(\cdot,\cdot,\cdot)$ is the Gaussian process defined in Theorem \ref{thm1}.
\end{corollary}

\section{Consistency and asymptotic local power}

We investigate the consistency and asymptotic local power properties of test statistics $CvM_n$ and $KS_n$ based on $S_n(w,y,z)$ under $\text{H}_1$ and under a sequence of local alternatives converging to $\text{H}_0$ at a parametric rate $n^{-1/2}$.

The asymptotic behavior of $S_n(w,y,z)$ under $\text{H}_1$ is stated in the next theorem.

\begin{theorem}\label{thm2}
    Suppose \eqref{markov} and Assumptions A1 - A7 in Appendix \ref{assumptions} hold.. Then under the alternative, for each $(w,y,z) \in \W \times \mathbb{R}^{d_y} \times \mathbb{R}^{d_z}$,
    \begin{equation*}
        n^{-1/2}S_n(w,y,z) \xrightarrow{P} \E\big[\varphi(W_t, w)\big(F_{Y,Z|W}(y,z|W_t)-F_{Y|W}(y|W_t)F_{Z|W}(z|W_t)\big)f_W(W_t)\big].
    \end{equation*}
\end{theorem}

Since $\E\big[\varphi(W_t, w)\big(F_{Y,Z|W}(y,z|W_t)-F_{Y|W}(y|W_t)F_{Z|W}(z|W_t)\big)f_W(W_t)\big]$ in a set with a positive Lebesgue measure under $\text{H}_1$, test statistics $CvM_n$ and $KS_n$ will diverge to infinity and have asymptotic power one against $\text{H}_1$.

To investigate the asymptotic local power properties of the tests, we introduce the following sequence of local alternatives,
\begin{equation}
\text{H}_{1n}: F_{Y,Z|W}(y,z|W_t)=F_{Y|W}(y|W_t)F_{Z|W}(z|W_t)+n^{-1/2}\Delta(W_t,y,z)\,\,\,\text{a.s.}\,\,\,\forall(y,z)\in\mathbb{R}^{d_y+d_z}, \label{alternative}
\end{equation}
where $\Delta(\cdot,\cdot,\cdot)$ is a non-constant measurable function, satisfying $\Delta(W_t,y,z)\neq 0$ a.s. for some $(y,z)$ and $\Delta(W_t,\infty,z)=\Delta(W_t,y,\infty)=\Delta(W_t,-\infty,z)=\Delta(W_t,y,-\infty)=0$ a.s. to deliver a valid conditional CDF $F_{Y,Z|W}(y,z|W_t)$ in \eqref{alternative}. The type of local alternatives in \eqref{alternative} is widely used in studying the asymptotic local power properties of tests based on empirical processes.

In \eqref{alternative}, the term $n^{-1/2}\Delta(W_t,y,z)$ characterizes the departure of the conditional joint CDF from the product of conditional marginal CDFs. In particular, $\Delta(W_t,y,z)$ specifies the direction of the departure, while $n^{-1/2}$ specifies the rate at which the difference between $F_{Y,Z|W}(y,z|W_t)$ and $F_{Y|W}(y|W_t)F_{Z|W}(z|W_t)$ shrinks to zero. Note that $n^{-1/2}$ is the fastest possible rate found in testing conditional independence.

To derive the local power result, we need the extra assumption i.e. Assumption A8 in Appendix \ref{assumptions}. The next theorem states the asymptotic behavior of $S_n(w,y,z)$ under $\text{H}_{1n}$.

\begin{theorem}\label{thm3}
    Suppose Assumption A1 - A8 in Appendix \ref{assumptions} hold. Then under the sequence of local alternatives in \eqref{alternative}, 
    \begin{equation*}
        S_n(\cdot,\cdot,\cdot)\rightsquigarrow S_{\infty}(\cdot,\cdot,\cdot)+G(\cdot,\cdot,\cdot),
    \end{equation*}
    where $S_{\infty}(\cdot,\cdot,\cdot)$ is the Gaussian process defined in Theorem \ref{thm1}.
\end{theorem}

Provided that the shift function $G(w,y,z):=\int^{w}_{-\infty}\Delta(\bar{w},y,z)f^2_W(\bar{w})\,d\bar{w}\neq 0$ in a set with a positive Lebesgue measure under $\text{H}_{1n}$, test statistics $CvM_n$ and $KS_n$ based on $S_n(x,y,z)$ will have non-trivial local power against $\text{H}_{1n}$ converging to $\text{H}_0$ at a parametric rate $n^{-1/2}$, the best rate known in testing conditional independence.

\begin{remark}
    For local power analysis, tests based on empirical processes can achieve a parametric rate $n^{-1/2}$, a rate much faster than those obtained from smoothing-based nonparametric tests. For example, \cite{su2008nonparametric} test only has power against local alternatives at a rate $n^{-1/2}h^{-d/4}$ with $d=d_w+d_y+d_z$, while tests proposed by \cite{su2007consistent} and \cite{bouezmarni2012nonparametric} have power against local alternatives at a rate $n^{-1/2}h^{-(d_w+d_z)/4}$. \cite{wang2018characteristic} test is the only one that has a better rate, which can detect local alternatives at a rate $n^{-1/2}h^{-d_w/4}$, though still slower than $n^{-1/2}$. Nonetheless, tests based on local smoothing are able to detect high frequency local alternatives considered by \cite{rosenblatt1975quadratic}, while our tests may not detect such type of local alternatives.
\end{remark}

The limiting distributions of $CvM_n$ and $KS_n$ under $\text{H}_{1n}$ is stated in the next result, which is a direct consequence of the continuous mapping theorem and Theorem \ref{thm3}.

\begin{corollary}\label{cor2}
    Suppose \eqref{markov} and Assumption A1 - A8 in Appendix \ref{assumptions} hold. Then under the sequence of local alternatives in \eqref{alternative},
    \begin{align*}
        CvM_n \rightsquigarrow \int_{\W \times \mathbb{R}^{d_y} \times \mathbb{R}^{d_z}}\big(S_{\infty}(w,y,z)+G(w,y,z)\big)^2\,dF_{W,Y,Z}(w,y,z),
    \end{align*}
    \begin{equation*}
        KS_n \rightsquigarrow \sup_{(w,y,z) \in \W \times \mathbb{R}^{d_y} \times \mathbb{R}^{d_z}}\left\vert S_{\infty}(w,y,z)+G(w,y,z)\right\vert,
    \end{equation*}
    where $S_\infty(\cdot,\cdot,\cdot)$ and $G(\cdot,\cdot,\cdot)$ are defined in Theorem \ref{thm3}.
\end{corollary}

Corollary \ref{cor2} implies that under $\text{H}_{1n}$ in \eqref{alternative}, the limiting distributions of $CvM_n$ and $KS_n$ are no longer the same as in Corollary 1 and they shift in a non-trivial way. Consequently, our tests have non-trivial power against the sequence of local alternatives in \eqref{alternative} converging to the null at a  parametric rate. 

\section{Bootstrap}\label{boot}
Since the asymptotic null distributions of $CvM_n$ and $KS_n$ depend on the underlying data generating process due to the complicated covariance kernel of the limiting process $S_\infty(w,y,z)$ in Theorem \ref{thm1}, it is difficult to tabulate the critical values for our tests. We propose a bootstrap procedure, which is in the spirit of the multiplier bootstrap suggested by \cite{delgado2001significance}. This bootstrap takes full advantage of the asymptotic theory in Theorem 1, and is also easy to implement as it does not have to compute new nonparametric estimates at each bootstrap replication.  

Let $\widehat\phi_t(y)=1(Y_{t}\leq y)-\widehat F_{Y|W}(y|W_t)$, $\widehat\epsilon_t(z)=1(Z_{t}\leq z)-\widehat F_{Z|W}(z|W_t)$, and $\widehat e_t(w,y,z) = \varphi(W_t, w) \widehat\phi_t(y)\widehat\epsilon_t(z)\widehat f_W(W_t)$, where $\widehat{F}_{Y|W}(y|W_t)$ is defined with $1(Y_s\leq y)$ replacing $1(Z_s\leq z)$ in $\widehat{F}_{Z|W}(z|W_t)$. The bootstrap empirical process of $S_{n}(w,y,z)$ is given by
\begin{align*}
    S_{n}^{\ast}(w,y,z)=\frac{1}{\sqrt n}\sum_{t=1}^n\widehat e_t(w,y,z)v_t,\label{rstar}
\end{align*}
where $\{v_t\}_{t=1}^n$ is a sequence of i.i.d. random variables with zero mean, unit variance, bounded support, and is independent of $\{(W^\top_t,Y^\top_t,Z^\top_t)^\top\}_{t=1}^n$. One popular choice due to \cite{mammen1993bootstrap} is the i.i.d. Bernoulli variates with probability masses given by
\begin{equation*}
    \pr \left(v_t=\frac{1-\sqrt{5}}{2}\right)=\frac{1+\sqrt{5}}{2\sqrt{5}}\,\,\,\text{and}\,\,\, \pr \left(v_t=\frac{1+\sqrt{5}}{2}\right)=\frac{-1+\sqrt{5}}{2\sqrt{5}}.
\end{equation*}
See also \cite{delgado2001significance} and \cite{escanciano2006generalized} for applications of this choice.

In the next theorem the asymptotic validity the bootstrap procedure is justified formally. We show that the bootstrapped process $S^\ast_{n}(\cdot,\cdot,\cdot)$ converges weakly to the Gaussian process $S_{\infty}(\cdot,\cdot,\cdot)$ in Theorem \ref{thm1}. As a result, we have convergence in distributions of the bootstrapped test statistics $CvM^\ast_n$ and $KS^\ast_n$, which are simply constructed by replacing $S_{n}(w,y,z)$ with $S^\ast_{n}(w,y,z)$ in \eqref{cvmn} and \eqref{ksn}, respectively.

\begin{theorem}\label{thm4}
    Suppose \eqref{markov} and Assumption A1 - A8 in Appendix \ref{assumptions} hold. Then, under the null, under the alternative, or under the sequence of local alternatives in \eqref{alternative},
    \begin{equation*}
        S_{n}^{\ast}(\cdot,\cdot,\cdot)\underset{\ast}{\overset{P}{\rightarrow}}S_{\infty}(\cdot,\cdot,\cdot),
    \end{equation*}
    where $S_{\infty}(\cdot,\cdot,\cdot)$ is the Gaussian process defined in Theorem \ref{thm1}, and $\underset{\ast}{\overset{P}{\rightarrow}}$ denotes weak convergence in probability under the bootstrap law, i.e., conditional on $\{(W^\top_t,Y^\top_t,Z^\top_t)^\top\}_{t=1}^n$. In addition, $CvM_n^\ast\underset{\ast}{\rightsquigarrow}CvM_\infty$ and $KS_n^\ast\underset{\ast}{\rightsquigarrow}KS_\infty$ with $CvM_\infty$ and $KS_\infty$ defined in Corollary \ref{cor1}.
\end{theorem}

Theorem \ref{thm4} implies that the limiting behavior of $S_n(w,y,z)$ can be approximated by that of $S^{\ast}_n(w,y,z)$. Thus, the bootstrap assisted tests $CvM_n$ and $KS_n$ have a correct asymptotic level, are consistent against $\text{H}_1$, and are able to detect local alternatives \eqref{alternative} converging to the null at a parametric rate. In practice, we can obtain the critical values of $CvM_n$ (and similarly for $KS_n$) as accurately as desired by the following algorithm:\\
\indent \textbf{Step 1.} Compute $S_n(W_t,Y_t,Z_t)$, and get $CvM_n$\\
\indent \textbf{Step 2.} Generate $\{v_t\}_{t=1}^n$ independently, compute $S_{n}^{\ast}(W_t,Y_t,Z_t)$, and get $CvM_n^{\ast}$.\\
\indent \textbf{Step 3.} Repeat \textbf{Step 2} $B$ times to have $\{CvM_{n,b}^{\ast}\}_{b=1}^B$, and compute its empirical $(1-\alpha)$-th sample quantile $CvM_{n}^{\ast\alpha}$ or bootstrapped $p$-value $p_n^{\ast }=B^{-1}\sum_{b=1}^{B}1\left( CvM_{n,b}^{\ast }\geq CvM_n\right)$. Rejects $\text{H}_0$ at the significance level $\alpha$ if $CvM_n>CvM_{n}^{\ast\alpha}$ or if $p_n^{\ast}<\alpha$.

\begin{remark}
    The above bootstrap assisted procedure applies when \eqref{markov} holds. If \eqref{markov} is violated, our proposed tests may suffer size distortion and power loss. However, under the general dependence structure in the data, it is possible to extend the block bootstrap (e.g., \cite{buhlmann1994blockwise}) to our context, the finite sample performance of which will be investigated through simulations in the next section. 
\end{remark}

\section{Monte Carlo simulations}\label{mc}
We carry out a set of Monte Carlo simulations to examine the finite sample performance of the proposed test statistics $CvM_n$ and $KS_n$. To examine the size performance, we consider the following four data generating processes (DGPs):
\begin{align*}
\text{(S1):} \quad Y_t &=\varepsilon_{1,t}, Z_t=\varepsilon_{2,t}, X_t=\varepsilon_{3,t}.\\
\text{(S2):} \quad Y_t &=0.5Y_{t-1}+\varepsilon_{1,t}.\\
%\text{(S3):} \quad Y_t &=\sqrt{h_t}\varepsilon_{1,t}, h_t=0.01+0.5Y_{t-1}^2\\
\text{(S3):} \quad Y_t &=0.5Y_{t-1}\exp{(-0.5Y^2_{t-1})}+\varepsilon_{1,t}.\\
\text{(S4):} \quad Y_t &=\sqrt{h_{1,t}}\varepsilon_{1,t}, h_{1,t}=0.01+0.9h_{1,t-1}+0.05Y_{t-1}^2,\\
Z_t &=\sqrt{h_{2,t}}\varepsilon_{2,t}, h_{2,t}=0.01+0.9h_{2,t-1}+0.05Z_{t-1}^2.
\end{align*}
To examine the power performance, the following seven DGPs are considered:
\begin{align*}
\text{(P1):} \quad Y_t &=0.5Y_{t-1}+0.5Z_{t-1}+\varepsilon_{1,t}.\\
\text{(P2):} \quad Y_t &=0.5Y_{t-1}+0.5Z_{t-1}^2+\varepsilon_{1,t}.\\
\text{(P3):} \quad Y_t &=0.5Y_{t-1}Z_{t-1}+\varepsilon_{1,t}.\\
\text{(P4):} \quad Y_t &=0.3+0.2\log(h_t)+\sqrt{h_{t}}\varepsilon_{1,t}, h_{t}=0.01+0.5Y_{t-1}^2+0.3Z_{t-1}^2.\\
\text{(P5):} \quad Y_t &=0.5Y_{t-1}+0.5Z_{t-1}\varepsilon_{1,t}.\\
\text{(P6):} \quad Y_t &=\sqrt{h_{t}}\varepsilon_{1,t}, h_{t}=0.01+0.5Y_{t-1}^2+0.25Z_{t-1}^2.\\
\text{(P7):} \quad Y_t &=\sqrt{h_{1,t}}\varepsilon_{1,t}, h_{1,t}=0.01+0.1h_{1,t-1}+0.4Y_{t-1}^2+0.5Z_{t-1}^2.\\
Z_t &=\sqrt{h_{2,t}}\varepsilon_{2,t}, h_{2,t}=0.01+0.9h_{2,t-1}+0.05Z_{t-1}^2.
\end{align*}
Here and below, $\varepsilon_{1,t}$, $\varepsilon_{2,t}$ and $\varepsilon_{3,t}$ are i.i.d. $N(0,1)$ and mutually independent, and $Z_t$ in (S2)-(S3) and (P1)-(P6) follows an AR(1) model: $Z_t =0.5Z_{t-1}+\varepsilon_{2,t}$. These DGPs cover a wide range of both linear and nonlinear time series processes. For (S1), $(Y_t,Z_t,X_t)$ is simply an i.i.d. sequence; we test $Y_t\bot Z_t|X_t$ such that $W_t=X_t$. From (S2) to (P7), we test $Y_t\bot Z_{t-1}|Y_{t-1}$ such that  $W_t=Y_{t-1}$; that is, we are interested in checking whether $Z_{t-1}$ has any predictive ability in explaining $Y_t$ (in mean, variance, or higher moments) after controlling the first lag of $Y_t$. Note that (S4) and (P7) do not satisfy \eqref{markov}.

For each DGP, we first generate $n+500$ observations and then discard the first 500 observations to minimize effects of initial values. The number of Monte Carlo simulations is $2000$ and the bootstrap critical values are obtained from $B=1000$ bootstrap replications. Four sample sizes, $n=100$, $200$, $400$ and $800$, are considered. We only report results for the nominal level of 5\% and results for other levels are available upon request. We choose the standard normal density $K(x)=(2\pi)^{-1/2}\exp{(-x^2/2)}$ as our kernel. Bandwidth of the form $h=c n^{-1/3.5}$ is used and results with $c=0.5$, $1.0$ and $1.5$ are reported to check the sensitivity of our tests to different bandwidths. How to choose $h$ to maximize the performance in our testing framework is beyond the scope of this paper and is left for future research. In addition to the fixed bandwidth $h$, in the supplemental appendix we provide more simulation results with a data-driven one $h=1.06\cdot Std(X_t)\cdot n^{-1/3.5}$.

Table \eqref{cvm_mb} and \eqref{ks_mb} report respectively the empirical rejection rates of $CvM_n$ and $KS_n$ under (S1)-(P7) using critical values obtained with the multiplier bootstrap method proposed in Section \ref{boot}. Both $CvM_n$ and $KS_n$ have acceptable empirical sizes in the cases of moderate sample sizes for (S1)-(S3), with $KS_n$ having more accurate sizes than $CvM_n$. When \eqref{markov} fails in (S4), $CvM_n$ exhibits a severe size distortion for $n=100$ and $n=200$, while $KS_n$ has less distortion; when sample size increases to 800, the two tests show improved sizes especially for $c=0.5$. For the empirical power performance, when sample size is as small as $n=100$, both $CvM_n$ and $KS_n$ are not very powerful against (P2)-(P7), with (P7) the most difficult DGP to detect. They gain power rapidly as sample size increases. Note that in (P7) \eqref{markov} fails, which might be the reason for low power in small samples. In summary, our simulation results indicate that $KS_n$ preserves sizes better than $CvM_n$, while $CvM_n$ is more powerful than $KS_n$; in addition, whether \eqref{markov} holds will affect the performance. The simulation results also show that the empirical sizes and powers of $CvM_n$ and $KS_n$ are somehow sensitive to bandwidth choices in very small samples; a general pattern is that larger $c$ tends to deliver a higher power but produces a bigger size distortion. When $n=800$, the overall performance is satisfactory.

\begin{center}
------------------------------------------

Tables \ref{cvm_mb} \& \ref{ks_mb} about here

------------------------------------------
\end{center}

To investigate if we have improved performance when we take into account the general dependence structure in the data, we also study the following block multiplier bootstrap:
\begin{equation*}
S_n^{\ast block}(w,y,z)=\frac{1}{\sqrt n}\sum_{t=1}^{n-L+1}\zeta_t\sum_{s=t}^{t+L-1}\widehat e_t(w,y,z),
\end{equation*}
where $\{\zeta_t\}_{t=1}^{n-L+1}$ are i.i.d. $N(0,L^{-1})$ and independent of the sample $\{(W_t^\top,Y_t^\top,Z_t^\top)^\top\}_{t=1}^n$, with $L:=L_n$ the block length diverging to infinity at a slower rate than $n$ as $n\to\infty$. Note that when $L=1$, $S_n^{\ast block}(w,y,z)$ reduces to the multiplier bootstrap introduced in Section \ref{boot}, the validity of which replies on the restriction \eqref{markov}. 

Table \eqref{cvm_bmb} and \eqref{ks_bmb} report respectively the empirical rejection rates of $CvM^{block}_n$ and $KS^{block}_n$ based on $S_n^{\ast block}(w,y,z)$. It is expected that $CvM^{block}_n$ and $KS^{block}_n$ should work when \eqref{markov} does not hold and consequently the dependence structure in the data should not be ignored. In implementing the block bootstrap, we have used $L=\lfloor an^{1/4} \rfloor$ with $a=1$, $2$ and $4$ to check the sensitivity of $CvM^{block}_n$ and $KS^{block}_n$ with respect to different block lengths. To save space, we only report the results for bandwidth $h=cn^{-1/3.5}$ with $c=1.0$. The results in Table \eqref{cvm_bmb} and \eqref{ks_bmb} are based on 1000 simulations and 200 bootstrap replications. It is observed that $CvM^{block}_n$ and $KS^{block}_n$ have less distorted empirical sizes especially for (S4) than $CvM_n$ and $KS_n$ do. At the same time, $CvM^{block}_n$ and $KS^{block}_n$ are not as powerful as $CvM_n$ and $KS_n$ even for (P7). Like before, $KS^{block}_n$ preserves sizes better than $CvM^{block}_n$, while $CvM^{block}_n$ is more powerful than $KS^{block}_n$. Lastly, for given $h$, the tests also depend on the choice of $L$; smaller $L$ tends to deliver a higher power but produces oversized tests.

\begin{center}
------------------------------------------

Tables \ref{cvm_bmb} \& \ref{ks_bmb} about here

------------------------------------------
\end{center}

\section{An empirical study}
We examine whether there exists nonlinear predictability of equity risk premium using variance risk premium. The variance risk premium is defined as the difference between the risk-neutral and objective expectations of realized variance, where the risk-neutral expectation of variance is measured as the end-of-month Volatility Index-squared de-annualized and the realized variance is the sum of squared 5-minute log returns of the S\&P 500 index over the month. 

There is a rich literature on the predictive power of variance risk premium for the aggregate stock market returns, bond returns or exchange rate returns. For example, Bollerslev et al. (2009) first discover that variance risk premium is able to explain a nontrivial fraction of the time series variation in post 1990 aggregate stock market returns, with high (low) premia predicting high (low) future returns; Wang et al. (2013) find the empirical evidence suggesting that the firm-level variance risk premium has a prominent explanatory power for credit spreads in the presence of market- and firm-level control variables; by defining a ``global'' variance risk premium, Bollerslev et al. (2013) uncover stronger predictability of aggregate stock market returns using variance risk premium across countries; while Della Corte et al. (2013) investigate the predictive information content in foreign exchange volatility risk premia for exchange rate returns and find that a portfolio that sells currencies with high insurance costs and buys currencies with low insurance costs generates sizeable out-of-sample returns and Sharpe ratios.

We use monthly aggregate S\&P 500 composite index over the period January 1996 to September 2008. Our empirical analysis is based on the logarithmic return on the S\&P 500 in excess of the 3-month T-bill rate. Let $RP_{t+\tau}$ be the risk premium $\tau$ months ahead and $VRP_{t}$ be the variance risk premium at time $t$. In this empirical study, we take $\tau=$1, 3, 6, and 9 months. We shall examine if the variance risk premium $VRP_t$ explains in a linear or nonlinear way the risk premium $RP_{t+\tau}$ given the information $RP_{t}$, which is equivalent to stating whether $VRP_t$ Granger causes $RP_t$ by setting the lag order to $\tau$. To test for the presence of (nonlinear) predictability of $VRP_t$, we consider to test 
\begin{equation*}
\text{H}_0: \pr\left\{F(RP_{t+\tau},VRP_{t}|RP_{t})=F(RP_{t+\tau}|RP_{t})F(VRP_t|RP_{t})\right\}=1
\end{equation*}
against $\text{H}_1: \pr\left\{F(RP_{t+\tau},VRP_{t}|RP_{t})=F(RP_{t+\tau}|RP_{t})F(VRP_t|RP_{t})\right\}<1$. That is, for a given horizon $\tau$, we test the conditional independence of $RP_{t+\tau}$ and $VRP_t$ given $RP_{t}$, i.e. $RP_{t+\tau}\bot VRP_{t}|RP_{t}$. 

For the purpose of comparison, we also perform the popular linear causality analysis in the literature. To this end, we consider the following linear regression model:
\begin{equation}
RP_{t+\tau}=\mu_{\tau}+\beta_{\tau}RP_{t}+\alpha_{\tau}VRP_{t}+\varepsilon_{t+\tau}. \label{lin}
\end{equation}
The hypothesis of interest is that VRP does not Granger cause RP for $\tau$ months ahead in a linear way, i.e. testing the null hypothesis $\text{H}_0:\alpha_{\tau}=0$ against the alternative
hypothesis $\text{H}_1:\alpha_{\tau}\neq 0$. To test $\text{H}_0$, standard $t$-statistic given by $t_{\hat{\alpha}_{\tau}}=\hat{\alpha}_{\tau}/\hat{\sigma}_{\hat{\alpha}_{\tau}}$ will be calculated, where $\hat{\alpha}_{\tau}$ is the least squares estimator of $\alpha_{\tau}$ and $\hat{\sigma}_{\hat{\alpha}_{\tau}}$ is the estimator of its standard error $\sigma_{\hat{\alpha}_{\tau}}$. Moreover, to avoid the impact of possible dependence in the residual terms $\hat{\varepsilon}_{t+\tau}$ on our inference, $\hat{\sigma}_{\hat{\alpha}_{\tau}}$ is calculated using the commonly used heteroscedasticity autocorrelation consistent (HAC) robust variance estimator suggested by Newey and West (1987).

Table \ref{empirical} reports the testing results for Granger causality (e.g. nonlinear predictability) from variance risk premium to risk premium, at four different horizons, using our proposed tests $CvM_n$ and $KS_n$ as well as the linear test. To check the robustness of our empirical findings, we also include the block multiplier tests $CvM_n^{block}$ and $KS_n^{block}$, where the block length is set to be $L=\lfloor an^{1/4} \rfloor$ with $a=2$. Results for $a=1$ and $4$ are similar and hence are omitted. The implementation of all bootstrap based tests is as demonstrated in the Monte Carlo simulations part with the number of bootstrap replications $B=10,000$. We have obtained encouraging findings from Table \ref{empirical} that are relevant to both empirical and theoretical studies of variance risk premium as a suitable predictor for risk premium. Specifically, results from linear regression based test in \eqref{lin} clearly fail to reject the null hypothesis of no linear predictability for one to six months (short-run) horizons; they only show weak evidence of linear predictability until at the nine months (long-run) horizon at the 5\% significance level. On the other hand, using our tests, we find convincing evidence that risk premium can be well predicted using variance risk premium at both mid-run and long-run horizons. Testing results are not sensitive to the bandwidth choices for six and nine months horizon, nor does the block type tests. For the one month horizon, we may conclude that variance risk premium has no predictive power, while the findings for three months horizon tend to indicate the existence of predictability. Overall, we find that there is a very high degree of predictability at horizons more than one-month which could be attributed to the nonlinear predictive effect. Our empirical evidence also indicates that caution is needed when interpreting results based on the linear regression.

\begin{center}
------------------------------------------

Table \ref{empirical} about here

------------------------------------------
\end{center}

\section{Conclusion}

This paper proposes new consistent nonparametric tests of conditional independence for time series data based on the empirical process method. The asymptotic properties of the proposed tests under the null, the alternative, and the sequence of local alternatives are investigated. To implement the test in practice, a multiplier bootstrap procedure is suggested and its asymptotic validity is formally justified. The test can be applied to testing for conditional independence in a wide variety of nonparametric models. Using the proposed test, we also study whether there exists some nonlinear predictability of equity risk premium using variance risk premium.

\bibliographystyle{chicago}     % Chicago style, author-year citations
\bibliography{ref} 

\newpage
\appendix
\renewcommand{\appendixname}{Appendix~\Alph{section}}
\section{Assumptions}\label{assumptions}

Denote $d: = d_w + d_y + d_z$.

\begin{itemize}
    
    \item[A1] $\{(X_t^{\top}, Y_t^{\top}, Z_t^{\top})^{\top}\}$ is a strictly stationary strong mixing process with mixing coefficients $\alpha(s)$ such that $\sum_{s = 1}^{\infty} s^{2m - 1} \alpha(s)^{\eta / (2m + \eta)} < \infty$ for some $m \in \mathbb{N}_+$ and $\eta > 0$ such that
    \begin{equation*}
        m \geq \frac{d + \sqrt{d^2 + 4 \eta d}}{4} \vee \frac{(d + \eta - 1) + \sqrt{(d + \eta - 1)^2 + 4 (d - 2) \eta}}{4}.
    \end{equation*}
    
    \item[A2] $\varphi(\cdot, \cdot)$ is uniformly bounded on the support of $W_t$ and $\W$, and $\varphi(w', w)$ is continuous about $w' \in \W$. Besides, there exists some measurable function $C_{\varphi}(\cdot)$ satisfying $\| C_{\varphi}\|_2 < \infty$,
    \begin{equation*}
        \big| \varphi (W, w) - \varphi (W, w')\big| \leq C_{\varphi}(W) |w - w'|^{\nu}
    \end{equation*}
    with arbitrary $\nu \in (0, 1]$. 
    
    \item[A3] $K(\cdot)$ is a product kernel of $k(\cdot)$ such that $K(u)=\prod_{j=1}^{d_w}k(u_j)$, $\int u^iK(u)\,du=\delta_{0i}$ for $i=0, 1, \cdots, l-1$ and $\int u^lK(u)\,du\neq 0$, where $k(\cdot)$ is a bounded, symmetric univariate kernel function and $\delta_{ij}$ is the delta function taking value one when $i=j$ and zero otherwise. 
    
    \item[A4] As $n\to\infty$, (i) $h\rightarrow 0$; (ii) $nh^{2d_w} / \log^{\xi} n \rightarrow \infty$ and $nh^{2l}\rightarrow 0$ for some positive $\xi > 0$.
    
    \item[A5] For any $m \in \{1, \ldots, n \}$ and $\{i_1, \ldots i_m \} \subset \{1, \ldots, n \}$, the joint density for $(W_{i_1}^{\top}, \ldots, W_{i_m}^{\top})^{\top}$ is bounded.
    
    \item[A6] The density of $Y$ and $Z$ exist and is bounded.
    
    \item[A7] $F_{Y|W}(y | W)$ satisfy the following local Holder continuity: for any $y, y' \in \mathbb{R}^{d_y}$, there exist a distribution function depending on $y, y'$ such that
    \begin{equation*}
        \big| F_{Y | W} (y | W) - F_{Y | W} (y' | W)\big| \leq C_{F_{Y | W}} (W) \big| G_{y, y'} (y) - G_{y, y'} (y') \big|^{\bar{\nu}}
    \end{equation*}
    for some $\bar{\nu} > 0$ satisfying 
    \begin{equation*}
        \bar{\nu} > 4 \left( 1 + \frac{2}{\eta}\right) \left( \frac{d_y}{d_y + 1} + \frac{d_z}{d_z + 1} \right)
    \end{equation*}
    with $\| C_{F_{Y |W}}(W)\|_2 < \infty$. The same requirement is imposed on $F_{Z|W}(z | W)$.
    
    \item[A8] (i) $\Delta(w, y, z)$ is uniformly bounded and is continuously differentiable with respect to $w$ up to order $l$, and continuously differentiable with respect to $y$ and $z$ (order $1$ is enough); (ii) $n^{-1} \sum_{t = 1}^n \varphi (W_t, w) \Delta (W_t, y, z) f_W (W_t) \xrightarrow{\text{as}} G(w, y, z) := \E [\varphi(W_1, w) \Delta (W_1, y, z) f_W(W_1)]$ uniformly in $(w, y, z) \in \W \times \mathbb{R}^{d_y} \times \mathbb{R}^{d_z}$.
\end{itemize}

\noindent \textbf{Remark on A1:} \textit{Assumption A1 says that the higher dimension of the data is, the stronger mixing condition we require!} 

\noindent \textbf{Remark on A2:} \textit{Assumption A2 says that we can use a very strong Holder property in the bounded space to control the exponential increase in the covering numbers cased by unbounded Euclidean space!} 

\noindent \textbf{Remark on A7:} \textit{Assumption A7 is not a strange assumption. As one can see, if we ignore the d.f. $G_{y, y'}$, Assumption A7 becomes:
\begin{equation*}
    \big| F_{Y | W} (y | W) - F_{Y | W} (y' | W)\big| \leq C_{F_{Y | W}} (W) | y - y' |^{\bar{\nu}},
\end{equation*}
which is a very common assumption in various literature. However, since $F_{Y|W}$ only value at $[0, 1]$, the above assumption is meaningless when the difference between $y$ and $y'$ larger than $1$. To avoid this problem, we introduce the function $G_{y, y'} : \mathbb{R}^{d_y} \rightarrow [0, 1]$. Since $G_{y, y'}$ depends on $y$ and $y'$, it is easy to see, the relationship that $| F_{Y | W} (y | W) - F_{Y | W} (y' | W)| \leq C_{F_{Y | W}} (W) | G_{y, y'} (y) - G_{y, y'} (y')|^{\bar{\nu}}$ can be easily satisfied if we didn't ask further restrictions on $G_{y, y'}$. However, in order to calculate the covering (or bracketing) number of the function class 
\begin{equation*}
    \mathcal{F}_{y, w} := \big\{ F_{Y | W} (y | \cdot) : \W \rightarrow [0, 1] \, ; \, y \in \mathbb{R}^{d_y}\big\},
\end{equation*}
some restrictions must be imposed on $G_{y, y'}$. Here we choose it as the distribution function (d.f.), any alternative is also possible. For instance, we can assume $G_{y, y'}$ is monotone when $d_y = 1$.
}

\section{Technical Lemmas}\label{lemma}

This section mainly serves for some part in Lemma \ref{lem.C1} that says that the larger dimension is, the more strong mixing condition requires. As well as the important bracketing lemma for conditional distribution functions.

\begin{customlemma}{B1}\label{lem.B1}
    Let $\{ V_i\}_{i = 1}^n$ be a $v$-dimensional strong mixing process with mixing coefficient $\alpha (\cdot)$. Let $F_{i_1, \ldots, i_m}$ denote the distribution function of $(V_{i_1}, \ldots, V_{i_m})$. For any integer $m > 1$ and integers $(i_1, \ldots, i_m)$ such that $1 \leq i_1 < i_2 < \cdots < i_m$, let $\theta$ be a Borel measurable function such that $\max \big\{ \int |\theta(v_1, \ldots, v_m)|^{1 + \widetilde{\eta}} \, dF_{i_1, \ldots, i_j} \, dF_{i_j + 1, \ldots, i_m}, \, \int |\theta(v_1, \ldots, v_m)|^{1 + \widetilde{\eta}} \, dF_{i_1, \ldots, i_m} \big\} \leq M_n $ for some $ \widetilde{\eta} > 0$. Then $\big| \int \theta(v_1, \ldots, v_m) \, dF_{i_1, \ldots, i_m} -  \int \theta(v_1, \ldots, v_m) \, dF_{i_1, \ldots, i_j} \, dF_{i_j + 1, \ldots, i_m} \big| \leq 4 M_n^{1 / (1 + \widetilde{\eta})} \alpha (i_{j + 1} - i_j)^{\widetilde{\eta} / (1 + \widetilde{\eta})}$.
\end{customlemma}
\begin{proof}
    See Lemma 2.1 in \cite{sun1997limiting}.
\end{proof}

Now denote the collection of probability measures
\begin{equation*}
    \begin{aligned}
        \mathcal{P}_j^k := \bigg\{ P_j^k (V_{i_1}, \ldots, V_{i_k}) := \prod_{s = 1}^j P (\underline{V}_s) : & \underline{V}_s \subset \{ V_{i_1}, \ldots, V_{i_k}\}, \\
        & \bigcup_{s = 1}^j \underline{V}_s = \{ V_{i_1}, \ldots, V_{i_k} \}, \  \underline{V}_s \cap \underline{V}_t = \emptyset \text{ for all } s \neq t\bigg\},
    \end{aligned}
\end{equation*}
and we want to derive the property of the $U$-processes
\begin{equation*}
    U_n := \frac{1}{n^2} \sum_{1 \leq i_1 < i_2 \leq n} \phi(v_{i_1}, v_{i_2})
\end{equation*}
with $\int \phi(v_1, v) \, d F_{v_i} (v) = \int \phi(v, v_2) \, d F_{v_i} (v) = 0$ for all $i$. For any $m \in \mathbb{N}_+$, let $I_m = \{ i_1, \ldots, i_{4 m}\}$, and define
\begin{equation*}
    \begin{aligned}
        M_{n 1 s}^{(m)} := & \max _{1 \leq i_{2 k-1}<i_{2 k} \leq n, 1 \leq k \leq 2 m \atop \text { exactly } (4 m + 1) -s \text { indices in } I_{m} \text { are distinct }} \max_{1 \leq j \leq (4 m + 1) - s} \\
        & \max_{P_{j}^{(4 m + 1)-s} \in \mathcal{P}_{j}^{(4 m + 1) - s}} \int \big| M(i_1, \ldots, i_{4 m})\big|^{1 + \eta / (2 m) } \,  d P_j^{(4 m + 1) - s}, \quad  s = 1,2, \ldots, 2 m
    \end{aligned}
\end{equation*}
\begin{equation*}
    \begin{aligned}
        M_{n 2 s}^{(m)} := & \max _{1 \leq i_{2 k-1}<i_{2 k} \leq n, 1 \leq k \leq 2 m \atop \text { exactly } (4 m - 3) - s \text { indices in } I_{m} \text { are distinct }} \max_{1 \leq j \leq (4 m - 3) - s} \\
        & \max_{P_{j}^{(4 m - 3) - s} \in \mathcal{P}_{j}^{(4 m - 3) - s}} \bigg| \int M(i_1, \ldots, i_{4 m}) \,  d P_j^{(4 m - 3) - s} \bigg|, \quad  s = 1,2, \ldots, 2m - 1.
    \end{aligned}
\end{equation*}
with $M(i_1, \ldots, i_{4m}) = \prod_{j = 1}^{2m} \phi (v_{i_{2j - 1}}, v_{i_{2j}})$. 

The following lemma is a extension for existent literature: \cite{yoshihara1976limiting} and \cite{su2012conditional} show the case for $m = 1$ and $m = 2, 3$ respectively.
\begin{customlemma}{B2}\label{lem.B2}
    If $\sum_{s = 1}^{\infty} s^{2m - 1} \alpha (s)^{\eta / (2 m + \eta)} < \infty$ for some $m \in \mathbb{N}_+$ and $\eta > 0$, then
    \begin{equation*}
        \E U_n^{2 m} = O \bigg( \sum_{s = 1}^{2 m} n^{- (2m - 1) - s} \big( M_{n1s}^{(m)} \big)^{2 m / (2 m + \eta)} + \sum_{s = 1}^{2m - 1} n^{- (2m + 1) - s} M_{n 2 s}^{(m)} \bigg).
    \end{equation*}
\end{customlemma}
\begin{proof}
    In the summation
    \begin{equation}\label{lem.B2.sum}
        \E U_n^{2m} = n^{- 4m} \sum_{j = 1}^{2 m} \sum_{1 \leq i_{2j - 1} < i_{2j} \leq n} \phi(v_{i_1}, v_{i_2}) \cdots \phi(v_{i_{4m - 1}}, v_{i_{4m}}),
    \end{equation}
    there are $(4m - 1)$ cases, denoted by $s = 1, \ldots, 4m - 1$, in which are exactly $(4m + 1) - s$ distinct indices among $i_1, \ldots, i_{4m}$. Let $1\leq k_1 \leq \ldots \leq k_{4m} \leq n$ be the permutation of $i_1, \ldots, i_{4m}$ in non-decreasing order, and 
    \begin{equation*}
        H(k_1, \ldots, k_{4m}) := \phi (v_{i_1}, v_{i_2}) \cdots \phi (v_{i_{4m - 1}}, v_{i_{4m}}).
    \end{equation*}
    For the first kind of case, i.e. $1 \leq s \leq 2m$. We denote
    \begin{equation*}
        \underline{d} := \min_{k_i < k_j} \big( k_j - k_i\big), \qquad \overline{d} := \max_{k_i < k_j} \big( k_j - k_i\big) = \overline{d} (k_1^*, k_2^*).
    \end{equation*}
    By Lemma \ref{lem.B1} with $\widetilde{\eta} = \eta / (2m)$, and noting that there are exactly $(4 m + 1) - s$ distinct indices in case ($s$), we have
    \begin{equation*}
        \begin{aligned}
            \E U_{n (s)} & \leq 4 \big( M_{n 1 s}^{(m)}\big)^{\frac{2m}{2m + \eta}} \underbrace{\sum \quad \cdots \quad  \sum}_{\text{the number of summation is exactly } (4m + 1) - s} \alpha^{\frac{\eta}{2m + \eta}} \big(\underline{d}\big) \\
            & \leq 4 \big( M_{n 1 s}^{(m)}\big)^{\frac{2m}{2m + \eta}} \sum_{1 \leq k_1^* < k_2^* \leq n} \overline{d}^{4m - 1 - s} \alpha^{\frac{\eta}{2m + \eta}} \big(\underline{d}\big) \\
            & \lesssim \big( M_{n 1 s}^{(m)}\big)^{\frac{2m}{2m + \eta}} n^{2m + 1 - s} \sum_{j = 1}^n j^{2m - 1} \alpha^{\frac{\eta}{2m + \eta}} (j),
        \end{aligned}
    \end{equation*}
    where the last step we use the fact $\sum_{j = 1}^{\infty}j^{2m - 1} \alpha^{\frac{\eta}{2m + \eta}} (j)$ is finite. Therefore, we have 
    \begin{equation*}
        \E U_{n(s)} = O \Big( \big( M_{n 1 s}^{(m)}\big)^{\frac{2m}{2m + \eta}} n^{2m + 1 - s} \Big), \qquad 1 \leq s \leq 2m.
    \end{equation*}
    For the second kind of cases, i.e. $2m + 1 \leq s \leq 4m - 1$, the expectation can be calculated directly. Indeed, since the indices in pairs $(i_{2j  - 1}, i_{2j})$ are different, $(4m + 1) - s$ distinct ($s = 2m + 1, \ldots, 4m - 1$) indices means the first ($i_{2j - 1}$) or second ($i_{2j}$) must be all the same, and hence,
    \begin{equation*}
        \E U_{n(s)} \leq M_{n 2 s}^{(m)} \underbrace{\sum \quad \cdots \quad \sum}_{\text{the number of summation is exactly } (4m + 1) - s} 1 = O \big( n^{(4m + 1) - s} M_{n 2 s}^{(m)} \big)
    \end{equation*}
    for $2m +1 \leq s \leq 4 m - 1$. Then the result in the lemma follows.
\end{proof}

\noindent Write $a_n \lesssim b_n$ as there exists some positive constant $c$ free of $n$ satisfying $a_n \leq c b_n$ for any $n \in \mathbb{N}$, and denote $a_n \asymp b_n$ if $a_n \lesssim b_n$ and $b_n \lesssim a_n$. And for notation simplicity, we denote $\gamma = (w^{\top}, y^{\top}, z^{\top})^{\top} \in \mathbb{R}^{d}$ with $d = d_w + d_y + d_z$.

\begin{customlemma}{B3}\label{lem.B3}
    Suppose Assumption A7 in Appendix \ref{assumptions} holds and $(Y, W)$ follow the joint law $P$, then the bracketing number of the class
    \begin{equation*}
        \F := \big\{ F_{Y | W} (y | \cdot) : \W \rightarrow [0, 1] \, ; \, y \in \mathbb{R}^{d_y}\big\}
    \end{equation*}
    satisfies
    \begin{equation*}
        \log N_{[ \; ]} \big(\varepsilon, \mathcal{F}, L_2(P) \big) \lesssim \varepsilon^{- 2 d_y / (\bar{\nu} (d_y + 1))}
    \end{equation*}
    for any $\varepsilon \in (0, 1]$
\end{customlemma}

\begin{proof}
    When $\| C_{F_{Y|W}}\|_2 = 0$, the result is trivial. Now suppose $\| C_{F_{Y|W}}\|_2 \neq 0$. Note that 
    \begin{equation*}
        \F \subseteq \big\{ F_{Y | W} (y | \cdot) : \W \rightarrow [0, 1] \text{ satisfying Assumption A7 }; \, G_{y, y'} : \mathbb{R}^{d_y} \rightarrow [0, 1] \text{ is a d.f.}\big\} =: \overline{\F}.
    \end{equation*}
    Then from the Assumption A7 by using a little extension of Theorem 2.7.11 in \cite{van1996weak} which says $\F = \{ f_t : t \in T \}$ have property that $|f_s(x) - f_t(x)| \leq d(s, t)^{\bar{\nu}} F(x)$ for some metric $d$ and fixed function $F(x)$, then for any norm $\| \cdot\|$, 
    \begin{equation*}
        N_{[\;]} \big( 2 \varepsilon^{\bar{\nu}} \| F\| , \F, \| \cdot \| \big) \leq N (\varepsilon, T, d)
    \end{equation*}
    with taking the norm $\| \cdot \| = \| \cdot \|_2$, we have
    \begin{equation*}
        \begin{aligned}
            \log N_{[ \; ]} \big( \varepsilon , \F, L_2(P) \big) & \leq \log N_{[ \; ]} \big( \varepsilon, \overline{\F}, L_2(P) \big) \\
            & \leq \log N \big( (2^{-1}\varepsilon / \| C_{F_{Y|W}}\|_2 )^{1 / \bar{\nu}}, \{ G_{y, y'} : \mathbb{R}^{d_y} \rightarrow [0, 1] \text{ is a d.f.} \}, L_2(P)\big) \\
            & \lesssim \varepsilon^{- \frac{2d_y}{\bar{\nu} (d_y + 1)}}
        \end{aligned}
    \end{equation*}
    where the last $\lesssim$ comes from Example 1.6.4 in \cite{wellner2002empirical}.
\end{proof}

\section{Main Proofs}\label{proof}

%Denote $\N (\epsilon, \F, L_r(P))$ as the covering number of $\F$ with respect to $L_r(P)$-norm, i.e. it equals the smallest number $N$ for which there exist functions $f_{1}, \ldots, f_{N}$ in $\F$ and $b_{1}, \ldots, b_{N}$ with $\|b \|_{P, r} := \big( \int b^r \, dP \big)^{1 / r} \leq \epsilon$ for each $i$ such that: for each $f$ in $\F$ there exists an $i$ for which $|f-f_{i}| \leq b_{i}$ (the introducing of $b_i$ is mainly for measurability, but we ignore this problem in the following content; \cite{andrews1994introduction} call this bracketing number, but people usually call it covering number in convention). And 

Denote $\gamma := (w^{\top}, y^{\top}, z^{\top}) \in \mathbb{R}^d$.

\begin{customlemma}{C1}\label{lem.C1}
    Under Assumptions A1 - A7 in Appendix \ref{assumptions} and null hypothesis,
    \begin{align*}
        \sup_{(w,y,z) \in \W \times \mathbb{R}^{d_y} \times \mathbb{R}^{d_z} }\left|S_{n}(w,y,z)-\bar S_{n}(w,y,z)\right|=o_p(1),
    \end{align*}
    with
    \begin{align*}
        \bar S_{n}(w,y,z)=\frac{1}{\sqrt n}\sum_{t=1}^n \varphi(W_t, w) \phi_t(y)\epsilon_t(z)f_W(W_t) := \frac{1}{\sqrt n}\sum_{t=1}^ne_t(w,y,z),
    \end{align*} 
    where
    \begin{align*}
        \phi_t(y)=1(Y_{t}\leq y)-F_{Y|W}(y|W_t),
    \end{align*}
    and
    \begin{align*}
        \epsilon_t(z)=1(Z_{t}\leq z)-F_{Z|W}(z|W_t).
    \end{align*}
\end{customlemma} 

\begin{proof}
    First, we note that $\W$ is bounded, whereas the value space for $Y_t$ and $Z_t$ is the whole Euclidean space. So we consider the bounded space
    \begin{equation*}
        \B_n := \big\{ (y, z) : |(y, z)| \leq \ell_n \big\} \subseteq \R^{d_y + d_z}
    \end{equation*}
    with some slowly increasing sequence $\ell_n \uparrow \infty$ we will specify later. Now the set $\Gamma_n := \W \times \B_n$ is bounded and can be cover by $\Gamma_{nk} := \{ \gamma_{n k} : |\gamma - \gamma_{nk}| \leq \delta_n , \, k = 1, \ldots, N_n \}$ with $\delta_n = n^{-1 / 2} (\log \log n) ^{-\kappa}$ for some positive $\kappa > 0$ and $N_n \asymp \ell_n^{d_y + d_z} \delta_n^{-d}$
     \begin{equation}
        \sup_{\gamma \in \Gamma_n}|S_n(\gamma)|\leq\max_{1\leq k\leq N_n}|S_n(\gamma_{nk})|+\max_{1\leq k\leq N_n}\sup_{\gamma\in\Gamma_n}|S_n(\gamma)-S_n(\gamma_{nk})|. \label{b1}
    \end{equation}
    
    Define $\chi_{t} = (W^\top_t,Y^\top_{t},Z^\top_{t})^\top$. We can rewrite $S_n(\gamma)$ as a form of standard $U$-statistic. Introduce
    \begin{align*}
        & U_\gamma(\chi_{t},\chi_{s})=\frac{1}{2 h^{d_w}}K\left(\frac{W_t-W_s}{h}\right) \varphi(W_t, w) 1(Y_t\leq y)(1(Z_t\leq z)-1(Z_s\leq z))\\
        &\qquad \qquad +\frac{1}{2 h^{d_w}}K\left(\frac{W_t-W_s}{h}\right) \varphi(W_s, w) 1(Y_s\leq y)(1(Z_s\leq z)-1(Z_t\leq z))\\
        =& \frac{1}{2h^{d_w}}K\left(\frac{W_t-W_s}{h}\right)\big(\varphi(W_t, w) 1(Y_t\leq y) - \varphi(W_s, w) 1(Y_s\leq y)\big)\big(1(Z_t\leq z)-1(Z_s\leq z)\big).
    \end{align*}
    Therefore, we have
    \begin{equation*}
        S_{n}(\gamma)=\frac{n}{n-1}\frac{1}{n^{3/2}}\sum_{1\leq s < t\leq n}^n U_\gamma(\chi_{s},\chi_{t}).
    \end{equation*}
    We reply on the $U$-statistic theory to study the asymptotic behavior of the above $U$-statistic; that is, by Hoeffding decomposition, we have
    \begin{equation}
        S_n(\gamma)=S^{(1)}_n(\gamma)+S^{(2)}_n(\gamma) , \label{b2}
    \end{equation}
    where
    \begin{equation*}
        S^{(1)}_n(\gamma) = \frac{2}{\sqrt n}\sum_{s=1}^n  \E U_{\gamma} (\chi, \chi_t) \big|_{\chi = \chi_s} := \frac{2}{\sqrt{n}} \sum_{s = 1}^n U_{1 \gamma} (\chi_s),
    \end{equation*}
    and
    \begin{equation*}
        S^{(2)}_n(\gamma)=\frac{1}{n^{1/2}(n-1)} \sum_{1\leq s<t\leq n}^n \Big[U_\gamma(\chi_s,\chi_t) - U_{1 \gamma} (\chi_s) - U_{1 \gamma} (\chi_t) \Big],
    \end{equation*}
    where $F_\chi(\chi_t)$ denotes the distribution of $\chi_t$. It is straightforward to show that
    \begin{align*}
        S^{(1)}_n(\gamma)=&\frac{1}{\sqrt n}\sum_{s=1}^n\int\frac{1}{h^{d_w}}K\left(\frac{W_t-W_s}{h}\right)\big(\varphi(W_t, w) 1(Y_t\leq y) - \varphi(W_s, w) 1(Y_s\leq y) \big)\\
        &\times \big(1(Z_t\leq z)-1(Z_s\leq z) \big)\,dF_\chi(\chi_t)\\
        =&\frac{1}{\sqrt n}\sum_{s=1}^n\int\frac{1}{h^{d_w}}K\left(\frac{W_t-W_s}{h}\right) \varphi(W_t, w) 1(Y_t\leq y)1(Z_t\leq z)\,dF_\chi(\chi_t)\\
        &-\frac{1}{\sqrt n}\sum_{s=1}^n\int\frac{1}{h^{d_w}}K\left(\frac{W_t-W_s}{h}\right)\varphi(W_t, w) 1(Y_t\leq y) 1(Z_s\leq z) \,dF_\chi(\chi_t)\\
        &-\frac{1}{\sqrt n}\sum_{s=1}^n \varphi(W_s, w) 1(Y_s\leq y)\int\frac{1}{h^{d_w}}K\left(\frac{W_t-W_s}{h}\right)1(Z_t\leq z)\,dF_\chi(\chi_t)\\
        &+\frac{1}{\sqrt n}\sum_{s=1}^n \varphi(W_s, w) 1(Y_s\leq y)1(Z_s\leq z)\int\frac{1}{h^{d_w}}K\left(\frac{W_t-W_s}{h}\right)\,dF_\chi(\chi_t)\\                                   :=& \sum_{j=1}^4 A_{nj},
    \end{align*}
    where in each integral, we regard $\chi_t$ as nonrandom variable with on relation to $\chi_s$. Let us first calculate $A_{n1}$, which is equal to
    \begin{equation*}
        \begin{aligned}
            & A_{n1} = \frac{1}{\sqrt{n}} \sum_{s = 1}^n \int \frac{1}{h^{d_w}} K \left( \frac{w' - W_s}{h}\right) \varphi(w', w) 1(y' \leq y) 1(z' \leq z) \, d F(w', y', z') \\
            & = \frac{1}{\sqrt{n}} \sum_{s = 1}^n \frac{1}{h^{d_w}} \int K \left( \frac{w' - w}{h}\right) \varphi(w', w) 1(y' \leq y) 1(z' \leq z) f_W(w') f_{Y|W} (y' | w') f_{Z|W} (z' | w') \, d w' \, dy' \, dz' \\
            & = \frac{1}{\sqrt n}\sum_{s=1}^n\int K(u) \varphi(W_s + hu, w) 1(y'\leq y)1(z'\leq z)\\
            &\qquad \qquad \qquad \times f_{Y|W}(y'|W_s+hu)f_{Z|W}(z'|W_s+hu)f_W(W_s+hu)\,du\,dy'\,dz'\\
            & = \frac{1}{\sqrt n}\sum_{s=1}^n \varphi(W_s + hu, w) \int 1(y'\leq y)f_{Y|W}(y'|W_s)\,dy'\int 1(z'\leq z)f_{Z|W}(z'|W_s)\,dz'f_W(W_s) + O_p(\sqrt nh^l)\\
            & = \frac{1}{\sqrt n}\sum_{s=1}^n \varphi(W_s, w) F_{Y|W}(y|W_s)F_{Z|W}(z|W_s)f_W(W_s)+o_p(1),
        \end{aligned}
    \end{equation*}
    where we apply the differentiable property of $f_{Y | W} (y | w)$, $f_{Z | W} (z | w)$, and $f_W(w)$ in Assumption A2 and the higher order kernel in Assumption A3 and the bandwidth condition in Assumption A4, and use the fact $F_{YZ|W}(y,z|w)=F_{Y|W}(y|w)F_{Z|W}(z|w)$ under the null of conditional independence, so that $dF_\chi(\chi')=f_{Y|W}(y'|w') \, f_{Z|W}(z'|w') f_W(w') \, dw' \, dy' \, dz'$. Similarly, we can find that
    \begin{align*}
        A_{n2} = -\frac{1}{\sqrt n} \sum_{s=1}^n \varphi(W_s, w) F_{Y|W}(y|W_s)1(Z_s\leq z)f_W(W_s)+o_p(1),
    \end{align*}
    \begin{align*}
        A_{n3}=-\frac{1}{\sqrt n}\sum_{s=1}^n \varphi(W_s, w) 1(Y_s\leq y)F_{Z|W}(z|W_s)f_W(W_s)+o_p(1),
    \end{align*}
    and
    \begin{align*}
        A_{n4}=\frac{1}{\sqrt n}\sum_{s=1}^n \varphi(W_s, w) 1(Y_s\leq y)1(Z_s\leq z)f_W(W_s)+o_p(1).
    \end{align*}
    Consequently,
    \begin{equation}
        S^{(1)}_n(\gamma) = \frac{1}{\sqrt n}\sum_{s=1}^n \varphi(W_s, w) \big( 1(Y_s\leq y)-F_{Y|W}(y|W_s)\big) \big(1(Z_s\leq z)-F_{Z|W}(z|W_s) \big)f_W(W_s)+o_p(1). \label{b3}
    \end{equation}
    
    Denote $\psi_\gamma(\chi_{s},\chi_{t}) = U_\gamma(\chi_s, \chi_t) - U_{1 \gamma} (\chi_s) - U_{1 \gamma} (\chi_t)$. We next show that
    \begin{equation*}
        S^{(2)}_n(\gamma)=\frac{1}{\sqrt n(n-1)}\sum_{1\leq s<t\leq n}^n \psi_\gamma(\chi_{s},\chi_{t})=\frac{1}{\sqrt n(n-1)}\sum_{t=s}^{n-1}\sum_{t=s+1}^n\psi_\gamma(\chi_{s},\chi_{t})=o_p(1), 
    \end{equation*}
    uniformly in $(w,y,z)$. For the $U$-process $n^{-1 / 2} S_n^{(2)} (\gamma)$, define $M_{n 1 s}^{(m)}$ and $M_{n 2 s}^{(m)}$ as in Appendix \ref{lemma}. For $s = 1$, we have all the probability measure in $\mathcal{P}_j^{(4m + 1) - 1}$ is separable. Denote $\{ \chi_t^*\}_{t = 1}^n$ as the same series to $\{ \chi_t\}_{t = 1}^n$ except $\chi_s^* \perp \chi_t^*$ for $s \neq t$, then the integral in $M_{n 1 1}^{(m)}$ is 
    \begin{equation*}
        \begin{aligned}
            \int \big| M(i_1, \ldots, i_{4 m})\big|^{1 + \eta / (2 m)} \,  d P_j^{4 m} & = \prod_{j = 1}^{2m} \int \big|\psi_{\gamma} (\chi_{i_{2 j - 1}}, \chi_{i_{2j}}) \big|^{1 + \eta / (2m)} \, dP_{i_{2 j - 1}} \, dP_{i_{2 j}} \\
            & = \prod_{j = 1}^{2m} \E \big|\psi_{\gamma} (\chi_{i_{2 j - 1}}^*, \chi_{i_{2j}}^*) \big|^{1 + \eta / (2m)} \leq \prod_{j = 1}^{2m} \sqrt{\E \psi_{\gamma}^{2 + \eta / m} (\chi_{i_{2 j - 1}}^*, \chi_{i_{2j}}^*) }.
        \end{aligned}
    \end{equation*}
    Now, for the second expectation in the above inequality, 
    \begin{equation*}
        \E \psi_{\gamma}^{2 + \eta / m} (\chi_{s}^*, \chi_t^*) \lesssim \E U_{\gamma}^{2 + \eta / m} (\chi_{s}^*, \chi_t^*) + 2 \E U_{1 \gamma}^{2 + \eta / m} (\chi_{s}^*),
    \end{equation*}
    with
    \begin{equation*}
        \begin{aligned}
            \E U_{\gamma}^{2 + \eta / m} (\chi_{s}^*, \chi_t^*) & = \frac{1}{(2h^{d_w})^{2 + \eta / m}} \int du \int K^{2 + \eta / m} \left( \frac{u - v}{h}\right) \big( \varphi(u, w) 1 (Y_s \leq y) \\
            & \qquad \qquad - \varphi(v, w) 1 (Y_t \leq y) \big)^{2 + \eta / m} \big( 1 (Z_t \leq z) - 1 (Z_s \leq z) \big)^{2 + \eta / m} \, dv \\
            & \lesssim \frac{1}{h^{(2 + \eta / m)d_w}} \int du \int K^{2 + \eta / m} \left( \frac{u - v}{h}\right) \, dv \asymp h^{- \eta d_w / m},
        \end{aligned}
    \end{equation*}
    where the last step comes from the boundedness of the set $\W$, and 
    \begin{equation*}
        \E U_{1 \gamma}^{2 + \eta / 2} (\chi_{s}^*) = \E \Big [ \E \big[ U_{\gamma}^{2 + \eta / m} (\chi_{s}^*, \chi_t^*) \, | \, \chi_t^* \big] \Big] \asymp h^{- \eta d_w / m},
    \end{equation*}
    which implies $\E \psi_{\gamma}^{2 + \eta / m} (\chi_{s}^*, \chi_t^*) = O \big( h^{- \eta d_w / m} \big)$. Hence
    \begin{equation*}
        \int \big| M(i_1, \ldots, i_{4m})\big|^{1 + \eta / (2m)} \,  d P_j^{2m} \lesssim \prod_{j = 1}^{2m} O \big( h^{- \eta d_w / (2m)} \big) = O \big( h^{- \eta d_w} \big),
    \end{equation*}
    and $M_{n 11}^{(m)} = O \big( h^{- \eta d_w} \big)$. However, for other $M_{n 1 s}^{(m)}$ ($s = 2, \ldots, 2m$), the integrals in them cannot be totally separable. For fixed $2 \leq s \leq 2m$, we denote $\{ \mathscr{D}_{\ell} \}_{\ell = 1}^{(4 m + 1) - s}$ are subsets chosen from $\{ i_1, \ldots, i_{4m}\}$ arbitrarily, then $M_{n 1 s}^{(m)} \leq \max_{t < s} M_{n 1 t}^{(m)}$ as well as
    \begin{equation}\label{lem.C1.dep.M}
        \begin{aligned}
            M_{n 1 s}^{(m)} & \leq \max_{i_j, \, 1 \leq i_j \leq 4m} \max_{\mathscr{D}_{\ell}, \, 1 \leq \ell \leq (4m + 1) - s} \int \big| M(i_1, \ldots, i_{4m}) \big|^{1 + \eta / (2m)} dP_j^{(4 m + 1) - s} \\
            & \leq \max_{i_j, \, 1 \leq i_j \leq 4m} \max_{\mathscr{D}_{\ell}, \, 1 \leq \ell \leq (4m + 1) - s} \Bigg[ \prod_{|\mathscr{D}_{\ell_1}| = 1} \int \prod_{i_{2j - 1}, i_{2j} \notin \mathscr{D}_{\ell_2} } \psi_{\gamma}^{2 + \eta / m} (\chi_{i_{2j - 1}}, \chi_{i_{2j}}) \,  d P_{\mathscr{D}_{\ell_1}} \\
            & \qquad \qquad \qquad \qquad \qquad \times \prod_{|\mathscr{D}_{\ell_2}| \geq 2} \int \prod_{i_{2j - 1} \text{ or } i_{2j} \in \mathscr{D}_{\ell_2}} \psi_{\gamma}^{2 + \eta / m} (\chi_{i_{2j - 1}}, \chi_{i_{2j}}) \,  d P_{\mathscr{D}_{\ell_2}} \Bigg]^{1 / 2}.
        \end{aligned}
    \end{equation}
    For finding the rate of the above integral, we first tackle each sub-sub-integral the last part of the sub-integral. By changing of variables, it is not hard to obtain that
    \begin{equation*}
        \begin{aligned}
            & \int \prod_{i_{2j - 1} \text{ or } i_{2j} \in \mathscr{D}_{\ell_2}} \psi_{\gamma}^{2 + \eta / m} (\chi_{i_{2j - 1}}, \chi_{i_{2j}}) \,  d P_{\mathscr{D}_{\ell_2}} \\
            & \leq \frac{1}{\big( h^{|\mathscr{D}_{\ell_2}| d_w} \big) ^{2 + \eta / m}} \int \prod_{a = 1}^{|\mathscr{D}_{\ell_2}|} K^{2 + \eta / m} \left( \frac{u_{i_a} - W_{i_a^*}}{h} \right) f_{W_{i_1^*}, \ldots, W_{i_{|\mathscr{D}_{\ell_2}|}^*} } \big(u_{i_1}, \ldots, u_{i_{|\mathscr{D}_{\ell_2}|}} \big) \, du \\
            & = \frac{1}{\big( h^{|\mathscr{D}_{\ell_2}| d_w} \big) ^{2 + \eta / m}} \times O \big( h^{|\mathscr{D}_{\ell_2}| d_w} \big) = O \big( h^{- (1 + \eta / m) |\mathscr{D}_{\ell_2}| d_w} \big),
        \end{aligned}
    \end{equation*}
    where 
    \begin{equation*}
        \bigcup_{a = 1}^{|\mathscr{D}_{\ell_2}|} \big\{ (i_a, i_a^* )\big\} = \bigcup_{i_{2j - 1} \text{ or } i_{2j} \in \mathscr{D}_{\ell_2}} \big\{ (i_{2j - 1}, i_{2j})\big\},
    \end{equation*}
    and the first "$=$" is due to the boundedness of $\W$ as well as Assumption A5. On the other hand, the order of the first sub-integral in \eqref{lem.C1.dep.M} is obvious due to the calculation of $M_{n 1 1}^{(m)}$ given the correct order of the second sub-integral:
    \begin{equation*}
        \begin{aligned}
            & \prod_{|\mathscr{D}_{\ell_1}| = 1} \int \prod_{i_{2j - 1}, i_{2j} \notin \mathscr{D}_{\ell_2} } \psi_{\gamma}^{2 + \eta / m} (\chi_{i_{2j - 1}}, \chi_{i_{2j}}) \,  d P_{\mathscr{D}_{\ell_1}} \\
            & = \prod_{i_{2j - 1}, i_{2j} \notin \mathscr{D}_{\ell_2} } \int \psi_{\gamma}^{2 + \eta / m} (\chi_{i_{2j - 1}}, \chi_{i_{2j}}) \,  d P_{i_{2j - 1}} \,  dP_{i_{2j}} = \prod_{i_{2j - 1}, i_{2j} \notin \mathscr{D}_{\ell_2} } \E \psi_{\gamma}^{2 + \eta / m} (\chi_{i_{2j - 1}}^*, \chi_{i_{2j}}^*) \\
            & = O \big(  h^{ - N (\mathscr{D}_{\ell_1}) \eta d_w / m } \big),
        \end{aligned}
    \end{equation*}
    where $ N (\mathscr{D}_{\ell_1})$ is the number of $\mathscr{D}_{\ell_1}$. Combining these results, we immediately get from \eqref{lem.C1.dep.M} that
    \begin{equation}\label{lem.C2.M.final}
        \begin{aligned}
            M_{n 1 s}^{(m)} & \leq \max_{\mathscr{D}_{\ell}, \, 1 \leq \ell \leq (4m + 1) - s} \sqrt{O \big(  h^{ - N (\mathscr{D}_{\ell_1}) \eta d_w / m } \big) \times O \big( h^{- N (\mathscr{D}_{\ell_2}) (1 + \eta / m) |\mathscr{D}_{\ell_2}| d_w}\big)} \\
            & = \max_{\mathscr{D}_{\ell}, \, 1 \leq \ell \leq (4m + 1) - s} O \Big(h^{- d_w / 2 \cdot \left( N(\mathscr{D}_{\ell_1}) \eta / m + (1 + \eta / m) N(\mathscr{D}_{\ell_2}) |\mathscr{D}_{\ell_2}| \right)} \Big),
        \end{aligned}
    \end{equation}
    where $N (\mathscr{D}_{\ell_1}) + N (\mathscr{D}_{\ell_2}) = (4 m + 1) - s$ and $N (\mathscr{D}_{\ell_1}) +  |\mathscr{D}_{\ell_2}| N (\mathscr{D}_{\ell_2}) = 4m$. This is a integer optimization problem, it is obvious that different $m$ may have different solutions. So we can only deal with the worst case. It is not hard to see what we want is to maximize the power coefficients in \eqref{lem.C2.M.final}, and hence the worst case in asymptotic sense is
    \begin{equation*}
        |\mathscr{D}_{\ell_2}| = s, \qquad N (\mathscr{D}_{\ell_1}) = 4m - s, \qquad N( \mathscr{D}_{\ell_2}) = 1.
    \end{equation*}
    Therefore, the order of $M_{n 1 s}^{(m)}$ is 
    \begin{equation*}
        M_{n 1 s}^{(m)} \leq \max_{1 \leq t < s} O \big( M_{n 1 t}^{(m)} \big) \vee O \Big(h^{- d_w (4 \eta + s) / 2} \Big) = O \Big(h^{- d_w (4 \eta + s) / 2} \Big), \qquad s = 2, 3, \ldots, 2m.
    \end{equation*}
    
    To determine the order of $M_{n 2 s}^{(m)}$, we first use the inequality,
    \begin{equation*}
        \begin{aligned}
            \bigg| \int M(i_1, \ldots, i_{4m}) \,  d P_j^{(4m - 3) - s} \bigg| & \leq \bigg| \int M_2 (\mathscr{I}_2)  \, d P_{2j}^{(4m - 3) - s}  \int \big| M_1(\mathscr{I}_1) \big| \,  d P_{1j}^{(4m - 3) - s} \bigg| \\
            & \leq \Bigg| \int M_2 (\mathscr{I}_2) \, d P_{2j}^{(4m - 3) - s} \sqrt{\int \big| M_1(\mathscr{I}_1) \big|^2 \,  d P_{1j}^{(4m - 3) - s}} \Bigg|,
        \end{aligned}
    \end{equation*}
    where the $\mathscr{I}_1 \cup \mathscr{I}_2 = \{ i_1, \ldots, i_{4m}\}$, $\mathscr{I}_1 \cap \mathscr{I}_2 = \emptyset$, and the following separable property guarantees,
    \begin{gather*}
        M(i_1, \ldots, i_{4m}) = M_1 (\mathscr{I}_1) \times M_2 (\mathscr{I}_2),\\
        P_{j}^{(4m - 3) - s} (\chi_{i_1}, \ldots, \chi_{i_8}) = P_{1j}^{(4m - 3) - s} \big( (\chi_{j})_{j \in \mathscr{I}_1} \big) \times P_{2j}^{(4m - 3) - s} \big( (\chi_{j})_{j \in \mathscr{I}_2} \big),
    \end{gather*}
    then by using the same notations and method as dealing with $M_{n 1 s}^{(m)}$, we can get $M_{n 2 s}^{(m)} \leq \max_{t < s} M_{n 2 t}^{(m)}$ (where we define $M_{n 2 0}^{(m)} = 1$), as well as 
    \begin{equation*}
        \begin{aligned}
            M_{n 2 s}^{(m)} & \leq \max_{i_j, \, 1 \leq i_j \leq 4m} \max_{\mathscr{D}_{\ell}, \, 1 \leq \ell \leq (4m - 3) - s} \Bigg| \prod_{|\mathscr{D}_{\ell_1}| = 1} \prod_{i_{2k - 1}, i_{2k} \notin \mathscr{D}_{\ell_2}} \int \psi_{\gamma} (\chi_{i_{2 k - 1}}, \chi_{i_{2k}}) \, d P_{\mathscr{D}_{\ell_1}} \\
            & \qquad \qquad \qquad \qquad \qquad \qquad \times \sqrt{\prod_{|\mathscr{D}_{\ell_2} | \geq 2} \int \prod_{i_{2 k - 1} \text{ or } i_{2k} \in \mathscr{D}_{\ell_2}} \psi_{\gamma}^2 (\chi_{i_{2 k - 1}}, \chi_{i_{2k}}) \, d P_{\mathscr{D}_{\ell_2}}} \, \Bigg| \\
            & \leq \max_{i_j, \, 1 \leq i_j \leq 4m} \max_{\mathscr{D}_{\ell}, \, 1 \leq \ell \leq (4m - 3) - s}  \bigg| O(1) \times \sqrt{\prod_{|\mathscr{D}_{\ell_2} | \geq 2} \frac{1}{(h^{|\mathscr{D}_{\ell_2} | d_w})^2} \cdot O (h^{|\mathscr{D}_{\ell_2} | d_w})} \bigg| \\
            & = O \big( h^{- \widetilde{N}(\mathscr{D}_{\ell_2}) |\mathscr{D}_{\ell_2}| d_w / 2}\big),
        \end{aligned}
    \end{equation*}
    where the number of $\mathscr{D}_{\ell_1}$ and $\mathscr{D}_{\ell_2}$ satisfies
    \begin{equation*}
        \widetilde{N}(\mathscr{D}_{\ell_1}) + \widetilde{N}(\mathscr{D}_{\ell_2}) = (4 m - 3) - s, \qquad \widetilde{N}(\mathscr{D}_{\ell_1}) + |\mathscr{D}_{\ell_2}| \widetilde{N}(\mathscr{D}_{\ell_2}) = 4m.
    \end{equation*}
    Similarly, it is easy to see the asymptotically worst case is
    \begin{equation*}
        |\mathscr{D}_{\ell_2}| = s, \qquad \widetilde{N}(\mathscr{D}_{\ell_1}) = 4 (m - 1) - s, \qquad \widetilde{N}(\mathscr{D}_{\ell_2}) = 1.
    \end{equation*}
    Therefore, the order of $M_{n 2 s}^{(m)}$ is
    \begin{equation*}
        M_{n 2 s}^{(m)} \leq \max_{0 \leq t < s} O \big( M_{n 1 t}^{(m)}\big) \vee O \big( h^{- (s + 4) d_w / 2}\big) = O \big( h^{- (s + 4) d_w / 2}\big), \qquad s = 1, \ldots, 2m - 1.
    \end{equation*}
    
    Now, from Lemma \ref{lem.B2}, 
    \begin{equation*}
        \E \big[ n^{- 1 / 2} S_n^{(2)} (\gamma) \big]^{2m} = O \bigg( \sum_{s = 1}^{2 m} n^{- (2m - 1) - s} \big( M_{n1s}^{(m)} \big)^{2 m / (2 m + \eta)} + \sum_{s = 1}^{2m - 1} n^{- (2m + 1) - s} M_{n 2 s}^{(m)} \bigg).
    \end{equation*}
    For the first summation,
    \begin{equation*}
        \begin{aligned}
            \sum_{s = 1}^{2 m} n^{- (2m - 1) - s} \big( M_{n1s}^{(m)} \big)^{2 m / (2 m + \eta)} & \asymp n^{- 2m} h^{- \eta d_w \cdot \frac{2m}{2m + \eta}} + \sum_{s = 2}^{2 m} n^{- (2 m - 1) - s} h^{- \frac{d_w (4 \eta + s)}{2} \cdot \frac{2m}{2m + \eta}} \\
            & = n^{- 2 m} h^{- \frac{2 m \eta d_w}{2m + \eta}} + n^{- 2m - 1} h^{- \frac{m d_w (4 \eta + 2)}{2m + \eta}}.
        \end{aligned}
    \end{equation*}
    where we note that for $s < t$,
    \begin{equation*}
        \frac{n^{- (2 m - 1) - s} h^{- \frac{d_w (4 \eta + s)}{2} \cdot \frac{2m}{2m + \eta}}}{n^{- (2 m - 1) - t} h^{- \frac{d_w (4 \eta + t)}{2} \cdot \frac{2m}{2m + \eta}}} = \Big( n h^{\frac{m d_w}{2m + \eta}}\Big)^{t - s} \gg \big( n h^{2d_w}\big)^{t - s} \rightarrow \infty.
    \end{equation*}
    Similarly, for the second summation,
    \begin{equation*}
        \sum_{s = 1}^{2m - 1} n^{- (2m + 1) - s} M_{n 2 s}^{(m)} \asymp \sum_{s = 1}^{2m - 1} n^{- (2m + 1) - s} h^{- (s + 4) d_w / 2} = n^{- 2(m + 1)} h^{- 5 d_w / 2}
    \end{equation*}
    with
    \begin{equation*}
        \frac{n^{- (2m + 1) - s} h^{- (s + 4) d_w / 2}}{n^{- (2m + 1) - t} h^{- (t + 4) d_w / 2}} = \big( n h^{d_w / 2} \big)^{t - s} \rightarrow \infty.
    \end{equation*}
    for any $s < t$. Therefore, we obtain that
    \begin{equation}\label{lem.C1.order}
        \E \big[ n^{- 1 / 2} S_n^{(2)} (\gamma) \big]^{2m} = O \Big( n^{- 2 m} h^{- \frac{2 m \eta d_w}{2m + \eta}} + n^{- 2m - 1} h^{- \frac{m d_w (4 \eta + 2)}{2m + \eta}} + n^{- 2(m + 1)} h^{- 5 d_w / 2} \Big).
    \end{equation}
    Furthermore, for any $\epsilon > 0$, by Markov inequality
    \begin{equation*}
        \begin{aligned}
            \pr \bigg( \max_{1 \leq k \leq N_n} \big|S_n^{(2)} (\gamma_{nk}) \big| \geq \epsilon \bigg) & \leq N_n \max_{1 \leq k \leq N_n} \pr \Big( \big| n^{- 1 / 2} S_n^{(2)} (\gamma_{nk}) \big| \geq n^{- 1 / 2} \epsilon \Big) \\
            & \leq \epsilon^{-4} N_n n^{m} \max_{1 \leq k \leq N_n} \E \big[ n^{- 1 / 2} S_n^{(2)} (\gamma_{nk}) \big]^{2m} \\
            & \asymp \epsilon^{-4} \big( A_1 + A_2 + A_3\big).
        \end{aligned}
    \end{equation*}
    From \eqref{lem.C1.order}, we can precisely give the order of $A_j$, $j = 1, 2, 3$. First, for $A_1$,
    \begin{equation*}
        A_1 = \ell_n^{d_y + d_z} n^{d / 2} (\log \log)^{d \kappa} n \cdot n^{- m} h^{- \frac{2 m \eta d_w}{2m + \eta}} = \frac{\ell_n^{d_y + d_z} (\log \log)^{\kappa d} n}{\log^{(m - d / 2) \xi} n} \left( \frac{n h^{2 d_w \cdot \frac{m \eta}{(2m + \eta) (m - d / 2)}}}{\log^{\xi} n} \right)^{- (m - d / 2)}.
    \end{equation*}
    Since by Assumption A1, we have
    \begin{equation*}
        m - d / 2 \geq 0, \qquad \frac{m \eta}{(2m + \eta) (m - d / 2)} \leq 1,
    \end{equation*}
    take 
    \begin{equation*}
        \left( \frac{\log^{(m - d / 2) \xi} n}{(\log \log)^{\kappa d} n} \right)^{1 / (d_y + d_z)} \gg \ell_n \uparrow \infty,
    \end{equation*}
    we have $A_1 = o(1)$. Similarly, take $\ell_n$ are above, for $A_2$ and $A_3$,
    \begin{gather*}
        A_2 = \frac{\ell_n^{d_y + d_z} (\log \log)^{\kappa d} n}{\log^{(m + 1 - d / 2) \xi} n} \left( \frac{n h^{2 d_w \cdot \frac{m (2 \eta + 1)}{(2m + \eta) (m + 1 - d / 2)}}}{\log^{\xi} n} \right)^{- (m + 1 - d / 2)} = o(1),\\
        A_3 = \frac{\ell_n^{d_y + d_z} (\log \log)^{\kappa d} n}{\log^{(m + 2 - d / 2) \xi} n} \left( \frac{n h^{2 d_w \cdot \frac{5}{4 (m + 2 - d / 2)}}}{\log^{\xi} n} \right)^{- (m + 2 - d / 2)} = o(1),
    \end{gather*}
    given
    \begin{equation*}
        \frac{m (2 \eta + 1)}{(2m + \eta) (m + 1 - d / 2)} \leq 1, \qquad \frac{5}{4 (m + 2 - d / 2)} \leq 1,
    \end{equation*}
    by Assumption A1. Thus
        \begin{equation}\label{b4}
            \max_{1 \leq k \leq N_n} \big|S_n^{(2)} (\gamma_{nk}) \big| = o_p(1).
        \end{equation}
    Next, for $\gamma_{nk} = (w_{nk},y_{nk},z_{nk})$, note that
        \begin{align*}
            &S_n(\gamma)-S_n(\gamma_{nk})\\
            =&\frac{1}{\sqrt n(n-1)h^{d_w}}\sum_{1 \leq s\neq t \leq n}K\left(\frac{W_t-W_s}{h}\right)\left[- \varphi(W_t, w) 1(Y_t\leq y)(1(Z_t\leq z)-1(Z_t\leq z_{nk}))\right.\\
            &+\varphi(W_t, w)1(Y_t\leq y)(1(Z_s\leq z)-1(Z_s\leq z_{nk}))\\
            &+\varphi(W_t, w)(1(Y_t\leq y)-1(Y_t\leq y_{nk}))(1(Z_t\leq z)-1(Z_t\leq z_{nk}))\\
            &-\varphi(W_t, w)(1(Y_t\leq y)-1(Y_t\leq y_{nk}))(1(Z_s\leq z)-1(Z_s\leq z_{nk}))\\
            &+(\varphi(W_t, w)-\varphi(W_t, w_{nk}))1(Y_t\leq y)(1(Z_t\leq z)-1(Z_t\leq z_{nk}))\\
            &-(\varphi(W_t, w)-\varphi(W_t, w_{nk}))1(Y_t\leq y)(1(Z_s\leq z)-1(Z_s\leq z_{nk}))\\
            &-(\varphi(W_t, w)-\varphi(W_t, w_{nk}))(1(Y_t\leq y)-1(Y_t\leq y_{nk}))(1(Z_t\leq z)-1(Z_t\leq z_{nk}))\\
            &+(\varphi(W_t, w)-\varphi(W_t, w_{nk}))(1(Y_t\leq y)-1(Y_t\leq y_{nk}))(1(Z_s\leq z)-1(Z_s\leq z_{nk}))\\
            &-\varphi(W_t, w)(1(Y_t\leq y)-1(Y_t\leq y_{nk}))(1(Z_t\leq z)-1(Z_s\leq z))\\
            &-(\varphi(W_t, w)-\varphi(W_t, w_{nk}))1(Y_t\leq y)(1(Z_t\leq z)-1(Z_s\leq z))\\
            &\left.+(\varphi(W_t, w)-\varphi(W_t, w_{nk}))(1(Y_t\leq y)-1(Y_t\leq y_{nk}))(1(Z_t\leq z)-1(Z_s\leq z))\right]\\
            :=&\sum_{j=1}^{11}B_{nj}(\gamma;\gamma_k).
        \end{align*}
    We shall deal with $B_{nj}(\gamma;\gamma_k)$ ($j=1,2\ldots,11$) separately. By the uniform boundedness of indicator function and kernel function, the absolute value of $B_{n1}(\gamma;\gamma_k)$ is no bigger than
    \begin{align*}
        &\frac{1}{\sqrt n}\sum_{t=1}^n \varphi(W_t, w) 1(Y_t\leq y)\left\vert 1(Z_t\leq z)-1(Z_t\leq z_{nk})\right\vert\frac{1}{(n-1)h^{d_w}}\sum_{s = 1, \, s\neq t}^n\left\vert K\left(\frac{W_t-W_s}{h}\right)\right\vert\\
        \leq & \frac{1}{\sqrt n}\sum_{t=1}^n\left\vert 1(Z_t\leq z)-1(Z_t\leq z_{nk})\right\vert O_p(1)\\
        =& o_p(\ell_n^{-1}) \times O_p(1) = o_p(\ell_n^{-1}),
    \end{align*}  
    by noting that the term $\frac{1}{(n-1)h^{d_w}}\sum_{s = 1, \, s\neq t}^n\left\vert K\left(\frac{W_t-W_s}{h}\right)\right\vert$ is $O_p(1)$ uniformly in $t$ by the LLN, and the term
    \begin{equation*}
        \frac{1}{\sqrt n}\sum_{t=1}^n |1(Z_t\leq z)-1(Z_t\leq z_{nk})| \sim \sqrt{n} \pr \big( z \leq Z_1 \leq z_{nk} \big) = O_p (\sqrt{n} \delta_n) = o_p(\ell_n^{-1})
    \end{equation*}
    by taking
    \begin{equation*}
        \left( \frac{\log^{(m - d / 2) \xi} n}{(\log \log)^{\kappa d} n} \right)^{1 / (d_y + d_z)} \wedge \log \log^{\kappa} n \gg \ell_n \uparrow \infty,
    \end{equation*}
    and the density of $Z_t$ is bounded in Assumption A6. It clearly follows that
    \begin{equation*}
        \max_{1\leq k\leq N_n}\sup_{\gamma \in \Gamma_{nk}}\left\vert B_{n1}(\gamma;\gamma_{nk})\right\vert = o_p(\ell_n^{-1}).
    \end{equation*}
    The absolute value of $B_{n2}(\gamma;\gamma_{nk})$ is no bigger than
    \begin{align*}
        &\frac{1}{\sqrt n}\sum_{s=1}^n\left\vert 1(Z_s\leq z)-1(Z_s\leq z_{nk})\right\vert\frac{1}{(n-1)h^{d_w}}\sum_{t = 1, \, t\neq s}^n \varphi(W_t, w) 1(Y_t\leq y)\left\vert K\left(\frac{W_t-W_s}{h}\right)\right\vert\\
        \lesssim & \frac{1}{\sqrt n}\sum_{s=1}^n\left\vert 1(Z_s\leq z)-1(Z_s\leq z_{nk})\right\vert\frac{1}{(n-1)h^{d_w}}\sum_{t = 1, \, t \neq s}^n\left\vert K\left(\frac{W_t-W_s}{h}\right)\right\vert\\
        =&o_p(\ell_n^{-1})\times O_p(1) = o_p(\ell_n^{-1}),
    \end{align*}
    leading to $\max_{1\leq k \leq N_n}\sup_{\gamma\in\Gamma_{nk}}\left\vert B_{n2}(\gamma;\gamma_k)\right\vert = o_p(\ell_n^{-1})$. Similarly, $\max_{1\leq k\leq N_n}\sup_{\gamma\in\Gamma_{nk}}\left\vert B_{nj}(\gamma;\gamma_{nk})\right\vert=o_p(1)$ for $j=3,4,\ldots,11$. Hence, $\max_{1\leq k\leq N_n}\sup_{\gamma\in\Gamma_{nk}}|S_n(\gamma)-S_n(\gamma_{nk})| = o_p(\ell_n^{-1})$. This result and \eqref{b1}, \eqref{b2}, \eqref{b3} and \eqref{b4} together imply that uniformly in $\gamma \in \Gamma_n$,
    \begin{equation}\label{lem.C2.last}
        \begin{aligned}
            & S_n(w,y,z) \\
            & = \frac{1}{\sqrt n}\sum_{s=1}^n1(W_s\leq w)\big((1(Y_s\leq y)-F_{Y|W}(y|W_s) \big) \big(1(Z_s\leq z)-F_{Z|W}(z|W_s)\big)f_W(W_s) + o_p(\ell_n^{-1})\\
            & = \bar{S}_n(w, y, z) + o_p(\ell_n^{-1})
        \end{aligned}
    \end{equation}
    Finally, we note that 
    \begin{equation*}
        \W \times \mathbb{R}^{d_y} \times \mathbb{R}^{d_z} = \bigcup_{n = 1}^{\infty} \Gamma_n = \lim_{n \rightarrow \infty} \Gamma_n,
    \end{equation*}
    take $n \rightarrow \infty$, we obtain uniform result on $\W \times \mathbb{R}^{d_y} \times \mathbb{R}^{d_z}$ in the lemma.
\end{proof}

Based on Lemma \ref{lem.C1}, now we can prove Theorem \ref{thm1}.
\vspace{2ex}

\textbf{Proof of Theorem \ref{thm1}:}
\begin{proof}
    By Lemma B.1, we have that, uniformly in $(w,y,z)$,
    \begin{equation*}
        S_n(w,y,z)=\frac{1}{\sqrt{n}}\sum_{t=1}^ne_t(w,y,z)+o_p(1) =: \bar S_n(w,y,z)+o_p(1).
    \end{equation*}
    It suffices to show that $\bar S_n(\cdot,\cdot,\cdot) \rightsquigarrow S_\infty(\cdot,\cdot,\cdot)$, where $S_\infty(\cdot,\cdot,\cdot)$ is defined in Theorem \ref{thm1}. Define the pseudometric $\rho_d$ on $\mathbb{R}^{d}$:
    \begin{equation*}
        \rho_d((w,y,z),(w',y',z')) := \{\E|e_t(w,y,z)-e_t(w',y',z')|^2\}^{1/2}
    \end{equation*}
    By Theorem 10.2 of \cite{pollard1990empirical}, this follows if we have (i) total boundedness of a pseudometric space $(\mathbb{R}^d,\rho_d)$, (ii) stochastic equicontinuity of $\{\bar S_n(w,y,z),n\geq 1\}$, and (iii) finite dimensional (fidi) convergence.
    
    Consider the class of functions 
    \begin{equation*}
        \mathcal{F} : = \{f_{(w,y,z)}:(w,y,z) \in \W \times \mathbb{R}^{d_y} \times \mathbb{R}^{d_z}\},
    \end{equation*}
    where $f_{(w,y,z)}:\mathbb{R}^{d_w}\times\mathbb{R}^{d_y}\times\mathbb{R}^{d_z}\rightarrow\mathbb{R}$ is defined by
    \begin{align*}
        f_{(w,y,z)}(W_t,Y_t,Z_t) = e_t(w,y,z) = \varphi(W_t, w) \phi_t(y)\epsilon_t(z)f_W(W_t).
    \end{align*}
    Consider $N(\varepsilon, \F, L_2(P))$ for $\varepsilon \in (0, 1]$, i.e. the covering number of $\F$. Since $f_W(\cdot)$ is bounded, it is equivalent to covering the no-necessarily bounded set $\Gamma = \W \times \Y \times \Z \subset \mathbb{R}^d$ by the cover
    \begin{equation*}
        \Gamma_k (\varepsilon) := \Gamma_k^{(\W)} (\varepsilon) \times \Gamma_k^{(\Y)} (\varepsilon) \times \Gamma_k^{(\Z)} (\varepsilon) = \big\{ \gamma_k = (w^{\top}, y^{\top}, z^{\top})^{\top} : k = 1, \ldots, N_{\varepsilon}\big\},
    \end{equation*}
    which satisfies for any $\gamma \in \Gamma$, there exists at least one $k$ such that
    \begin{equation*}
        \big\| \varphi(W_t, w) \phi_t(y) \epsilon_t (z) - \varphi(W_t, w) \phi_t(y) \epsilon_t (z) \big\|_2 \leq \varepsilon.
    \end{equation*}
    Obviously, the covering number satisfies
    \begin{equation*}
        N(\varepsilon, \F, L_2(P)) \asymp N_{\varepsilon} \leq N \big(\varepsilon, \F_w, L_2(P) \big) \times N \big(\varepsilon, \F_y, L_2(P) \big) \times N \big(\varepsilon, \F_z, L_2(P) \big),
    \end{equation*}
    where
    \begin{equation*}
        \F_w := \{ \varphi(W_t, w) : w \in \W\}, \quad \F_y := \{ \phi_t (y) : y \in \Y \}, \quad \F_z := \{ \epsilon_t (z) : z \in \Z \},
    \end{equation*}
    and the $P$ is the joint law for $(W_t^{\top}, Y_t^{\top}, Z_t^{\top})$. Since $\W$ is bounded, we immediately have
    \begin{equation*}
        N \big(\varepsilon, \F_w, L_2(P) \big) \lesssim \varepsilon^{- d_w / \nu}.
    \end{equation*}
    For $N(\varepsilon, \F_y, L_2(P))$, from Example 1.6.4 in \cite{wellner2002empirical}, 
    \begin{equation*}
        N\big( \varepsilon / 2, \big\{ 1(Y_t \leq y): y \in \mathbb{R}^{d_y}\big\}, L_2(P) \big) \lesssim \varepsilon^{- 2 d_y};
    \end{equation*}
    And from Lemma \ref{lem.B3}, we know that
    \begin{equation*}
        \begin{aligned}
            & \log N \big( \varepsilon / 2, \big\{ F_{Y | W} (y | \cdot) : \W \rightarrow [0, 1] \, ; \, y \in \mathbb{R}^{d_y}\big\}, L_2(P)\big) \\
            & \leq \log N_{[\;]} \big( \varepsilon, \big\{ F_{Y | W} (y | \cdot) : \W \rightarrow [0, 1] \, ; \, y \in \mathbb{R}^{d_y}\big\}, L_2(P)\big) \lesssim \varepsilon^{- \frac{2 d_y}{\bar{\nu} (d_y + 1)}}.
        \end{aligned}
    \end{equation*}
    Hence,
    \begin{equation*}
        \begin{aligned}
            & N \big( \varepsilon, \F_y, L_2(P)\big) \\
            & \leq N\big( \varepsilon / 2, \big\{ 1(Y_t \leq y): y \in \mathbb{R}^{d_y}\big\}, L_2(P) \big) \times N \big( \varepsilon / 2, \big\{ F_{Y | W} (y | \cdot) : \W \rightarrow [0, 1] \, ; \, y \in \mathbb{R}^{d_y}\big\}, L_2(P)\big) \\
            & \lesssim \varepsilon^{- 2d_y} \exp \left\{ \varepsilon^{- \frac{2 d_y}{\bar{\nu} (d_y + 1)}} \right\}
        \end{aligned}
    \end{equation*}
    Similarly, we have
    \begin{equation*}
        N \big( \varepsilon, \F_z, L_2(P)\big) \lesssim \varepsilon^{- 2d_z} \exp \left\{ \varepsilon^{- \frac{2 d_z}{\bar{\nu} (d_z + 1)}} \right\}.
    \end{equation*}
    As a consequence, for any $\varepsilon \in (0, 1]$, the covering number of the cover number of $\F$ satisfies
    \begin{equation*}
        N(\varepsilon, \F, L_2(P)) \lesssim \varepsilon^{- (\nu^{-1} d_w + 2 (d_y + d_z))} \cdot \exp \Big\{ \varepsilon^{ - \frac{2}{\bar{\nu}} \left( \frac{d_y}{d_y + 1} + \frac{d_z}{d_z + 1} \right)}\Big\}.
    \end{equation*}
    
    Consequently, 
    \begin{equation*}
        \int_0^1 \epsilon^{- \frac{\eta}{2 + \eta}}\log^2  \N \big(\epsilon, \F_1, L_r(P)\big) \, d\varepsilon \lesssim \int_0^1 \varepsilon^{- \left( \frac{\eta}{2 + \eta} + \frac{4}{\bar{\nu}} \left( \frac{d_y}{d_y + 1} + \frac{d_z}{d_z + 1} \right) \right)} \,  d \varepsilon < \infty,
    \end{equation*}
    with 
    \begin{equation*}
        \frac{\eta}{2 + \eta} + \frac{4}{\bar{\nu}} \left( \frac{d_y}{d_y + 1} + \frac{d_z}{d_z + 1} \right) < 1,
    \end{equation*}
    and $\sum_{j = 1}^{\infty} j^{2m - 1} \alpha(j)^{\frac{\eta}{2m + \eta}} < \infty$ by Assumption A1 and A7. It follows that the conditions are satisfied by a generalization of Theorem 2.2 of \cite{andrews1994introduction} (see p128 in \cite{andrews1994introduction}). The fidi convergence follows by the Cramer-Wold device and a central limit theorem for bounded random variables under strong mixing conditions; see Corollary 5.1 in \cite{hall1981rates}. We are left to demonstrate that the sample mean and covariance kernel converges to that of the limiting Gaussian process $S_\infty(\cdot,\cdot,\cdot)$ under the Markov property (2.1), i.e.,
    \begin{equation*}
        \begin{aligned}
            \E \big[\bar{S}_n(w,y,z) \big] & =\frac{1}{\sqrt{n}}  \sum_{t=1}^n \E \big[e_s(w,y,z)\big] \\
            & = \frac{1}{\sqrt{n}}  \sum_{t=1}^n \E \Big[ 1(W_t \leq w) f_W(W_t) \big( \E [1(Y_t \leq y) \, | \, W_t] - F_{Y|W} (y | W_t)\big) \\
            & \qquad \qquad \qquad \qquad \qquad \qquad \quad \times \big( \E [1(Z_t \leq z) \, | \, W_t] - F_{Z|W} (z | W_t)\big) \Big] = 0,
        \end{aligned}
    \end{equation*}
    and the covariance under IID case can be derived similarly. This completes the proof of the theorem.
\end{proof}

\textbf{Proof of Corollary \ref{cor1}:}

\begin{proof}
    By continuous mapping theorem, see e.g. Theorem 5.1 in \cite{billingsley2013convergence}, and the weak convergence of the process $S_n(w,y,z)$ in Theorem 1, we immediately have the convergence of $KS_n$ to $KS_\infty$ in distribution, e.g. \cite{delgado2007nonparametric}.

    For the limiting null distribution of $CvM_n$, note that
    \begin{align*}
        &|CvM_n-CvM_{\infty}|\\
        =&\left|\int S^2_n(w,y,z)\,dF_{n}(w,y,z)-\int S^2_{\infty}(w,y,z)\,dF_{W,Y,Z}(x,y,z)\right|\\
        \leq & \left|\int(S^2_n(w,y,z)-S^2_{\infty}(w,y,z))\,dF_{n}(w,y,z)\right|\\
        &+ \left|\int S^2_{\infty}(w,y,z)\,d(F_{n}(w,y,z)-F_{W,Y,Z}(w,y,z))\right|.
    \end{align*}
    The first term of the right-hand side of the above inequality is $o_p(1)$ by Theorem 1. Glivenko-Cantelli's Theorem yields that $\sup_{(w,y,z)}|F_n(w,y,z)-F_{W,Y,Z}(w,y,z)|=o(1)$ a.s. Then, taking into account that the trajectories of the limiting process $S_{\infty}(\cdot,\cdot,\cdot)$ are bounded and continuous almost surely and applying Helly-Bray Theorem (see p. 97 in \cite{rao1973linear}) to each of these trajectories, we obtain 
    \begin{equation*}
        \left|\int S^2_{\infty}(w,y,z)\,d(F_{n}(w,y,z)-F_{W,Y,Z}(w,y,z))\right|\rightarrow 0\,\,\,\text{a.s.}.
    \end{equation*}
    This concludes the proof of the corollary.
\end{proof}

\textbf{Proof of Theorem \ref{thm2}:}

\begin{proof}
    We rewrite
    \begin{align*}
        \frac{1}{\sqrt n} S_n(w,y,z)=&\frac{1}{n}\sum_{t=1}^n \varphi(W_t, w) 1(Y_t\leq y)(1(Z_t\leq z)-F_{Z|W}(z|W_t))f_W(W_t)\\
        &-\frac{1}{n}\sum_{t=1}^n \varphi(W_t, w) 1(Y_t\leq y)(\widehat F_{Z|W}(z|W_t)-F_{Z|W}(z|W_t))\widehat f_W(W_t)\\
        &+\frac{1}{n}\sum_{t=1}^n \varphi(W_t, w) 1(Y_t\leq y)(1(Z_t\leq z)-F_{Z|W}(z|W_t))(\widehat f_W(W_t)-f_W(W_t))\\
        :=&\frac{1}{\sqrt n} S_{n1}(w,y,z)+\frac{1}{\sqrt n} S_{n2}(w,y,z)+\frac{1}{\sqrt n} S_{n3}(w,y,z).
    \end{align*}
    From the consistency of the kernel estimators under Assumption A1 (since we at least have $m \geq 1$ hold in Assumption A1 hold, then \cite{su2012conditional} for example immediately gives the consistency) and boundedness assumption for $\varphi(\cdot, \cdot)$ and $\W$,
    \begin{equation*}
        \begin{aligned}
            \frac{1}{\sqrt{n}} S_{n2} (w, y, z) & = \frac{1}{n} \sum_{t = 1}^n \varphi(W_t, w) 1(Y_t \leq y) \cdot o_p(1) \cdot \big( f_W(W_t) + o_p(1)\big) \\
            & = o_p(1) \cdot \E \big[ \varphi(W_t, w) 1(Y_t \leq y) f_W(W_t) \big] = o_p(1).
        \end{aligned}
    \end{equation*}
    And similarly $n^{- 1 / 2} S_{n3} (w, y, z) = o_p(1)$. It follows that
    \begin{align*}
        \frac{1}{\sqrt n} S_n(w,y,z)=&\frac{1}{n}\sum_{t=1}^n \varphi(W_t, w) 1(Y_t\leq y)(1(Z_t\leq z)-F_{Z|W}(z|W_t))f_W(W_t)+o_p(1)\\
        =&\E\Big[\varphi(W_t, w) \big( (F_{Y,Z|W}(y,z|W_t)-F_{Y|W}(y|W_t)F_{Z|W}(z|W_t)\big)f_W(W_t)\Big]+o_p(1),
    \end{align*}
    where the last step is due to LLN and law of iterate expectations. Since
    \begin{equation*}
        \E \big[1(W_t\leq w)(F_{Y,Z|W}(y,z|W_t)-F_{Y|W}(y|W_t)F_{Z|W}(z|W_t))f_W(W_t) \big]\neq 0
    \end{equation*}
    in a set with positive Lebesgue measure, the test statistics based on $S_n(w,y,z)$, e.g. $CvM_n$ and $KS_n$, diverge to infinity under the alternative. In particular, $n^{-1}CvM_n$ (cf. the proof of Corollary \ref{cor1}) and $n^{-1/2}KS_n$ converge in probability to some positive values. This ends the proof of the theorem.
\end{proof}

\textbf{Proof of Theorem \ref{thm3}:}

\begin{proof}
    Similar to Lemma \ref{lem.C1}, in this case, we also have
        \begin{equation*}
            S_n(\gamma) = S_n^{(1)} (\gamma) + S_n^{(2)} (\gamma)
        \end{equation*}
    with $S_n^{(1)} (\gamma) = \sum_{j = 1}^4 A_{nj}$, where the definition of each symbol is exactly the same as that in Lemma \ref{lem.C1} until now. Now under $\text{H}_{1n}$, the term $A_{n1}$ is 
    \begin{equation*}
        \begin{aligned}
            A_{n1} & = \frac{1}{\sqrt{n} h^{d_w}} \sum_{s = 1}^n \int K \left( \frac{W_t - W_s}{h}\right) \varphi(W_t, w) 1 (Y_t \leq y) 1 (Z_t \leq z) \, d F_{\chi} (\chi_t) \\
            & = \frac{1}{\sqrt n}\sum_{s=1}^n \int \int \int  K(u) \varphi(W_s + hu, w)1(y'\leq y)1(z'\leq z)\\
            &\times \bigg[ f_{Y|W}(y'|W_s+hu)f_{Z|W}(z'|W_s+hu)+n^{-1/2}\frac{\partial^2\Delta(W_s,y',z')}{\partial y'\partial z'} \bigg] f_W(W_s+hu)\, du \, dy' \, dz'\\
            & =\frac{1}{\sqrt n}\sum_{s=1}^n \varphi(W_s, w) \int 1(y'\leq y)f_{Y|W}(y'|W_s)\,dy'\int 1(z'\leq z)f_{Z|W}(z'|W_s)\,dz'f_W(W_s)\\
            &+\frac{1}{n}\sum_{s=1}^n \varphi(W_s, w) \int \int  1(y'\leq y)1(z'\leq z)\frac{\partial^2\Delta(W_s,y',z')}{\partial y'\partial z'}\,dy'\,dz'f_W(W_s)+O_p(\sqrt nh^l)\\
            & = \frac{1}{\sqrt n}\sum_{s=1}^n\varphi(W_s, w) F_{Y|W}(y|W_s)F_{Z|W}(z|W_s)f_W(W_s)\\
            &+\frac{1}{n}\sum_{s=1}^n \varphi(W_s, w) \Delta(W_s,y,z)f_W(W_s)+o_p(1)\\
            & := \frac{1}{\sqrt n}\sum_{s=1}^n \varphi(W_s, w)F_{Y|W}(y|W_s)F_{Z|W}(z|W_s)f_W(W_s)+G(w,y,z)+o_p(1),
        \end{aligned}
    \end{equation*}
    where the first step is due to $f_{YZ|W}(y,z|W_t)=f_{Y|W}(y|W_t)f_{Z|W}(z|W_t)+n^{-1/2}\partial^2\Delta(W_t,y,z)/\partial y\partial z$ under $\text{H}_{1n}$, while the last step follows by Assumption A8 (ii). On the other hand, we can easily find that $A_{n2}$, $A_{n3}$ and $A_{n4}$ stay the same as in Lemma \ref{lem.C1}. As a result, following similar arguments of Lemma \ref{lem.C1}, we get
    \begin{align*}
        S_n(w,y,z)=&\frac{1}{\sqrt n}\sum_{s=1}^n \varphi(W_s, w) \big(1(Y_s\leq y)-F_{Y|W}(y|W_s)\big)\big(1(Z_s\leq z)-F_{Z|W}(z|W_s)\big)f_W(W_s)\\
        &+G(w,y,z)+o_p(1),
    \end{align*}
    uniformly in $(w,y,z) \in \W \times \mathbb{R}^{d_y} \times \mathbb{R}^{d_z}$ under the local alternatives $\text{H}_{1n}$. By the proof of Theorem \ref{thm1}, under the local alternatives $\text{H}_{1n}$,
    \begin{equation*}
        S_n(w,y,z) \rightsquigarrow S_{\infty}(\cdot,\cdot,\cdot)+G(\cdot,\cdot,\cdot),
    \end{equation*}
    which concludes the proof of the theorem.
\end{proof}

\textbf{Proof of Theorem \ref{thm4}:}

\begin{proof}
    We need to show that the process
    \begin{align*}
        S_n^\ast(w,y,z)=&\frac{1}{\sqrt n}\sum_{t=1}^n \varphi(W_t, w) (1(Y_t\leq y)-\widehat F_{Y|W}(y|W_t))(1(Z_t\leq z)-\widehat F_{Z|W}(z|W_t))\widehat f_W(W_t)v_t\\
        :=&\frac{1}{\sqrt n}\sum_{t=1}^n\widehat e_t(w,y,z)v_t
    \end{align*}
    (conditionally on the sample) has the same asymptotic fidis as the process $S_n(w,y,z)$ in probability, total boundedness of the pseudometric space $\big(\W \times \mathbb{R}^{d_y} \times \mathbb{R}^{d_z}, \rho_d \big)$ and stochastic equicontinuity of $\{S_n^\ast(w,y,z), n\geq 1\}$, both in probability. Recall that $\{v_t\}_{t=1}^n$ is a sequence of independent random variables with zero mean, unit variance and is independent of the original sample. 

    Define
    \begin{align*}
        \bar S_n^{\ast}(w,y,z)=&\frac{1}{\sqrt n}\sum_{t=1}^n \varphi(W_t, w) (1(Y_t\leq y)-F_{Y|W}(y|W_t))(1(Z_t\leq z)-F_{Z|W}(z|W_t))f_W(W_t)v_t\\
        :=&\frac{1}{\sqrt n}\sum_{t=1}^ne_t(w,y,z)v_t.
    \end{align*}
    It suffices to prove that, uniformly in $(w,y,z)$, the process $S_n^{\ast}(w,y,z)$ and the process $\bar S_n^{\ast}(w,y,z)$ are asymptotically equivalent, i.e.
    \begin{equation*}
        \sup_{(w,y,z) \in \W \times \mathbb{R}^{d_y} \times \mathbb{R}^{d_z}}\left|S_n^{\ast}(w,y,z)-\bar S_n^{\ast}(w,y,z)\right|=o_p(1).
    \end{equation*} 
    To show this, note that
    \begin{align*}
        &S_n^{\ast}(w,y,z)-\bar S_n^{\ast}(w,y,z)\\
        =&\frac{1}{\sqrt{n}}\sum_{t=1}^n \varphi(W_t, w) \big(1(Y_t\leq y)-\widehat{F}_{Y|W}(y|W_t)\big) \big(1(Z_t\leq z)-\widehat{F}_{Z|W}(z|W_t)\big)(\widehat{f}_W(W_t)-f_W(W_t))v_t\\
        -& \frac{1}{\sqrt{n}}\sum_{t=1}^n \varphi(W_t, w) \big(\widehat{F}_{Y|W}(y|W_t)-F_{Y|W}(y|W_t)\big)(1(Z_t\leq z)-F_{Z|W}(z|W_t))f_W(W_t)v_t\\
        -& \frac{1}{\sqrt{n}}\sum_{t=1}^n \varphi(W_t, w) \big(1(Y_t\leq y)-F_{Y|W}(y|W_t)\big)\big(\widehat{F}_{Z|W}(z|W_t)-F_{Z|W}(z|W_t)\big)f_W(W_t)v_t\\
        +& \frac{1}{\sqrt{n}}\sum_{t=1}^n \varphi(W_t, w) \big(\widehat{F}_{Y|W}(y|W_t)-F_{Y|W}(y|W_t)\big)\big(\widehat{F}_{Z|W}(z|W_t)-F_{Z|W}(z|W_t)\big)f_W(W_t)v_t\\
        :=& \sum_{j=1}^4 D_{nj}.
    \end{align*}
    
    Following the uniform consistency of the kernel estimators and together with the properties of $\{v_t\}_{t=1}^n$, it is easy to show that $D_{nj}=o_p(1)$ ($j=1,2,3,4$), uniformly in $(w,y,z) \in \W \times \mathbb{R}^{d_y} \times \mathbb{R}^{d_z}$.
    
    So, we only have to focus on $\bar S_n^\ast(w,y,z)$. The fidi convergence of $\bar S_n^\ast(w,y,z)$ follows from its expression and from the Cramer-Wold device in conjunction with \eqref{markov}. The total boundedness and stochastic equicontinuity of $\{\bar S_n^\ast(w,y,z),n\geq 1\}$ follow from the same steps as in Theorem \ref{thm1}. The proof is finished.
\end{proof}

\section{Simulation Tables}\label{tables}

\begin{table}[!htbp]
\caption{Empirical rejection rates of $CvM_n$ with $h=cn^{-1/3.5}$}\label{cvm_mb} 
\begin{equation*}
\begin{tabular}{@{}clllllllllll}
\hline\hline
 & \multicolumn{11}{c}{DGPs} \\ \hline
\multicolumn{1}{c}{$c$} & (S1) & (S2) & (S3) & (S4) & (P1) & 
(P2) & (P3) & (P4) & (P5)& (P6)& (P7)\\ \hline
\multicolumn{12}{c}{} \\ 
\multicolumn{12}{c}{$n=100$} \\ 
&  &  &  &  &  &  & \\ 
\multicolumn{1}{c}{$0.5$} & \multicolumn{1}{r}{0.052} & \multicolumn{1}{r}{0.058} & \multicolumn{1}{r}{0.064} & \multicolumn{1}{r}{0.073} & \multicolumn{1}{r}{0.966} & 
\multicolumn{1}{r}{0.297} & \multicolumn{1}{r}{0.136} & \multicolumn{1}{r}{0.215} & \multicolumn{1}{r}{0.140} & \multicolumn{1}{r}{0.127} & \multicolumn{1}{r}{0.096}\\ 
\multicolumn{1}{c}{$1.0$} & \multicolumn{1}{r}{0.062} & \multicolumn{1}{r}{0.074} & \multicolumn{1}{r}{0.077} & \multicolumn{1}{r}{0.088} & \multicolumn{1}{r}{0.982} & 
\multicolumn{1}{r}{0.372} & \multicolumn{1}{r}{0.165} & \multicolumn{1}{r}{0.245} & \multicolumn{1}{r}{0.214} & \multicolumn{1}{r}{0.150} & \multicolumn{1}{r}{0.127}\\ 
\multicolumn{1}{c}{$1.5$} & \multicolumn{1}{r}{0.077} & \multicolumn{1}{r}{0.086} & \multicolumn{1}{r}{0.088} & \multicolumn{1}{r}{0.118} & \multicolumn{1}{r}{0.987} & 
\multicolumn{1}{r}{0.399} & \multicolumn{1}{r}{0.206} & \multicolumn{1}{r}{0.281} & \multicolumn{1}{r}{0.282} & \multicolumn{1}{r}{0.200} & \multicolumn{1}{r}{0.142}\\ 
&  &  &  &  &  &  & \\ 
\multicolumn{12}{c}{$n=200$} \\ 
&  &  &  &  &  &  & \\ 
\multicolumn{1}{c}{$0.5$} & \multicolumn{1}{r}{0.051} & \multicolumn{1}{r}{0.053} & \multicolumn{1}{r}{0.045} & \multicolumn{1}{r}{0.059} & \multicolumn{1}{r}{1.000} & 
\multicolumn{1}{r}{0.854} & \multicolumn{1}{r}{0.345} & \multicolumn{1}{r}{0.542} & \multicolumn{1}{r}{0.327} & \multicolumn{1}{r}{0.210} & \multicolumn{1}{r}{0.127}\\ 
\multicolumn{1}{c}{$1.0$} & \multicolumn{1}{r}{0.063} & \multicolumn{1}{r}{0.053} & \multicolumn{1}{r}{0.063} & \multicolumn{1}{r}{0.082} & \multicolumn{1}{r}{1.000} & 
\multicolumn{1}{r}{0.902} & \multicolumn{1}{r}{0.400} & \multicolumn{1}{r}{0.607} & \multicolumn{1}{r}{0.424} & \multicolumn{1}{r}{0.242} & \multicolumn{1}{r}{0.177}\\  
\multicolumn{1}{c}{$1.5$} & \multicolumn{1}{r}{0.074} & \multicolumn{1}{r}{0.079} & \multicolumn{1}{r}{0.079} & \multicolumn{1}{r}{0.102} & \multicolumn{1}{r}{1.000} & 
\multicolumn{1}{r}{0.920} & \multicolumn{1}{r}{0.412} & \multicolumn{1}{r}{0.620} & \multicolumn{1}{r}{0.501} & \multicolumn{1}{r}{0.285} & \multicolumn{1}{r}{0.190}\\ 
&  &  &  &  &  &  & \\ 
\multicolumn{12}{c}{$n=400$} \\ 
&  &  &  &  &  &  & \\ 
\multicolumn{1}{c}{$0.5$} & \multicolumn{1}{r}{0.060} & \multicolumn{1}{r}{0.052} & \multicolumn{1}{r}{0.049} & \multicolumn{1}{r}{0.061} & \multicolumn{1}{r}{1.000} & 
\multicolumn{1}{r}{0.999} & \multicolumn{1}{r}{0.787} & \multicolumn{1}{r}{0.976} & \multicolumn{1}{r}{0.934} & \multicolumn{1}{r}{0.625} & \multicolumn{1}{r}{0.258}\\ 
\multicolumn{1}{c}{$1.0$} & \multicolumn{1}{r}{0.066} & \multicolumn{1}{r}{0.056} & \multicolumn{1}{r}{0.064} & \multicolumn{1}{r}{0.066} & \multicolumn{1}{r}{1.000} & 
\multicolumn{1}{r}{1.000} & \multicolumn{1}{r}{0.818} & \multicolumn{1}{r}{0.979} & \multicolumn{1}{r}{0.960} & \multicolumn{1}{r}{0.663} & \multicolumn{1}{r}{0.335}\\ 
\multicolumn{1}{c}{$1.5$} & \multicolumn{1}{r}{0.065} & \multicolumn{1}{r}{0.069} & \multicolumn{1}{r}{0.069} & \multicolumn{1}{r}{0.089} & \multicolumn{1}{r}{1.000} & 
\multicolumn{1}{r}{1.000} & \multicolumn{1}{r}{0.824} & \multicolumn{1}{r}{0.990} & \multicolumn{1}{r}{0.967} & \multicolumn{1}{r}{0.692} & \multicolumn{1}{r}{0.361}\\ 
&  &  &  &  &  &  & \\ 
\multicolumn{12}{c}{$n=800$} \\ 
&  &  &  &  &  &  & \\ 
\multicolumn{1}{c}{$0.5$} & \multicolumn{1}{r}{0.048} & \multicolumn{1}{r}{0.044} & \multicolumn{1}{r}{0.049} & \multicolumn{1}{r}{0.053} & \multicolumn{1}{r}{1.000} & 
\multicolumn{1}{r}{1.000} & \multicolumn{1}{r}{0.998} & \multicolumn{1}{r}{1.000} & \multicolumn{1}{r}{1.000} & \multicolumn{1}{r}{1.000} & \multicolumn{1}{r}{0.860}\\ 
\multicolumn{1}{c}{$1.0$} & \multicolumn{1}{r}{0.065} & \multicolumn{1}{r}{0.059} & \multicolumn{1}{r}{0.052} & \multicolumn{1}{r}{0.067} & \multicolumn{1}{r}{1.000} & 
\multicolumn{1}{r}{1.000} & \multicolumn{1}{r}{0.998} & \multicolumn{1}{r}{1.000} & \multicolumn{1}{r}{1.000} & \multicolumn{1}{r}{1.000} & \multicolumn{1}{r}{0.878}\\ 
\multicolumn{1}{c}{$1.5$} & \multicolumn{1}{r}{0.065} & \multicolumn{1}{r}{0.067} & \multicolumn{1}{r}{0.054} & \multicolumn{1}{r}{0.063} & \multicolumn{1}{r}{1.000} & 
\multicolumn{1}{r}{1.000} & \multicolumn{1}{r}{0.999} & \multicolumn{1}{r}{1.000} & \multicolumn{1}{r}{1.000} & \multicolumn{1}{r}{1.000} & \multicolumn{1}{r}{0.896}\\  \hline\hline
\end{tabular}
\end{equation*}
\end{table}

\begin{table}[!htbp]
\caption{Empirical rejection rates of $KS_n$ with $h=cn^{-1/3.5}$}\label{ks_mb} 
\begin{equation*}
\begin{tabular}{@{}clllllllllll}
\hline\hline
 & \multicolumn{11}{c}{DGPs} \\ \hline
\multicolumn{1}{c}{$c$} & (S1) & (S2) & (S3) & (S4) & (P1) & 
(P2) & (P3) & (P4) & (P5)& (P6)& (P7)\\ \hline
\multicolumn{12}{c}{} \\ 
\multicolumn{12}{c}{$n=100$} \\ 
&  &  &  &  &  &  & \\ 
\multicolumn{1}{c}{$0.5$} & \multicolumn{1}{r}{0.042} & \multicolumn{1}{r}{0.044} & \multicolumn{1}{r}{0.035} & \multicolumn{1}{r}{0.055} & \multicolumn{1}{r}{0.800} & 
\multicolumn{1}{r}{0.236} & \multicolumn{1}{r}{0.096} & \multicolumn{1}{r}{0.231} & \multicolumn{1}{r}{0.146} & \multicolumn{1}{r}{0.115} & \multicolumn{1}{r}{0.072}\\ 
\multicolumn{1}{c}{$1.0$} & \multicolumn{1}{r}{0.047} & \multicolumn{1}{r}{0.047} & \multicolumn{1}{r}{0.048} & \multicolumn{1}{r}{0.063} & \multicolumn{1}{r}{0.862} & 
\multicolumn{1}{r}{0.290} & \multicolumn{1}{r}{0.104} & \multicolumn{1}{r}{0.270} & \multicolumn{1}{r}{0.200} & \multicolumn{1}{r}{0.130} & \multicolumn{1}{r}{0.102}\\ 
\multicolumn{1}{c}{$1.5$} & \multicolumn{1}{r}{0.047} & \multicolumn{1}{r}{0.055} & \multicolumn{1}{r}{0.046} & \multicolumn{1}{r}{0.066} & \multicolumn{1}{r}{0.889} & 
\multicolumn{1}{r}{0.325} & \multicolumn{1}{r}{0.141} & \multicolumn{1}{r}{0.269} & \multicolumn{1}{r}{0.255} & \multicolumn{1}{r}{0.156} & \multicolumn{1}{r}{0.117}\\ 
&  &  &  &  &  &  & \\ 
\multicolumn{12}{c}{$n=200$} \\ 
&  &  &  &  &  &  & \\ 
\multicolumn{1}{c}{$0.5$} & \multicolumn{1}{r}{0.042} & \multicolumn{1}{r}{0.036} & \multicolumn{1}{r}{0.040} & \multicolumn{1}{r}{0.049} & \multicolumn{1}{r}{0.999} & 
\multicolumn{1}{r}{0.667} & \multicolumn{1}{r}{0.197} & \multicolumn{1}{r}{0.534} & \multicolumn{1}{r}{0.361} & \multicolumn{1}{r}{0.217} & \multicolumn{1}{r}{0.122}\\ 
\multicolumn{1}{c}{$1.0$} & \multicolumn{1}{r}{0.049} & \multicolumn{1}{r}{0.038} & \multicolumn{1}{r}{0.054} & \multicolumn{1}{r}{0.056} & \multicolumn{1}{r}{0.996} & 
\multicolumn{1}{r}{0.718} & \multicolumn{1}{r}{0.219} & \multicolumn{1}{r}{0.587} & \multicolumn{1}{r}{0.397} & \multicolumn{1}{r}{0.243} & \multicolumn{1}{r}{0.158}\\  
\multicolumn{1}{c}{$1.5$} & \multicolumn{1}{r}{0.051} & \multicolumn{1}{r}{0.052} & \multicolumn{1}{r}{0.059} & \multicolumn{1}{r}{0.072} & \multicolumn{1}{r}{0.997} & 
\multicolumn{1}{r}{0.730} & \multicolumn{1}{r}{0.232} & \multicolumn{1}{r}{0.587} & \multicolumn{1}{r}{0.459} & \multicolumn{1}{r}{0.274} & \multicolumn{1}{r}{0.169}\\ 
&  &  &  &  &  &  & \\ 
\multicolumn{12}{c}{$n=400$} \\ 
&  &  &  &  &  &  & \\ 
\multicolumn{1}{c}{$0.5$} & \multicolumn{1}{r}{0.047} & \multicolumn{1}{r}{0.043} & \multicolumn{1}{r}{0.042} & \multicolumn{1}{r}{0.051} & \multicolumn{1}{r}{1.000} & 
\multicolumn{1}{r}{0.986} & \multicolumn{1}{r}{0.462} & \multicolumn{1}{r}{0.938} & \multicolumn{1}{r}{0.779} & \multicolumn{1}{r}{0.510} & \multicolumn{1}{r}{0.237}\\ 
\multicolumn{1}{c}{$1.0$} & \multicolumn{1}{r}{0.055} & \multicolumn{1}{r}{0.044} & \multicolumn{1}{r}{0.051} & \multicolumn{1}{r}{0.056} & \multicolumn{1}{r}{1.000} & 
\multicolumn{1}{r}{0.989} & \multicolumn{1}{r}{0.467} & \multicolumn{1}{r}{0.944} & \multicolumn{1}{r}{0.816} & \multicolumn{1}{r}{0.527} & \multicolumn{1}{r}{0.275}\\ 
\multicolumn{1}{c}{$1.5$} & \multicolumn{1}{r}{0.055} & \multicolumn{1}{r}{0.051} & \multicolumn{1}{r}{0.054} & \multicolumn{1}{r}{0.073} & \multicolumn{1}{r}{1.000} & 
\multicolumn{1}{r}{0.997} & \multicolumn{1}{r}{0.501} & \multicolumn{1}{r}{0.958} & \multicolumn{1}{r}{0.839} & \multicolumn{1}{r}{0.545} & \multicolumn{1}{r}{0.289}\\ 
&  &  &  &  &  &  & \\ 
\multicolumn{12}{c}{$n=800$} \\ 
&  &  &  &  &  &  & \\ 
\multicolumn{1}{c}{$0.5$} & \multicolumn{1}{r}{0.052} & \multicolumn{1}{r}{0.051} & \multicolumn{1}{r}{0.042} & \multicolumn{1}{r}{0.055} & \multicolumn{1}{r}{1.000} & 
\multicolumn{1}{r}{1.000} & \multicolumn{1}{r}{0.856} & \multicolumn{1}{r}{1.000} & \multicolumn{1}{r}{0.999} & \multicolumn{1}{r}{0.936} & \multicolumn{1}{r}{0.575}\\ 
\multicolumn{1}{c}{$1.0$} & \multicolumn{1}{r}{0.058} & \multicolumn{1}{r}{0.052} & \multicolumn{1}{r}{0.043} & \multicolumn{1}{r}{0.063} & \multicolumn{1}{r}{1.000} & 
\multicolumn{1}{r}{1.000} & \multicolumn{1}{r}{0.873} & \multicolumn{1}{r}{1.000} & \multicolumn{1}{r}{0.995} & \multicolumn{1}{r}{0.954} & \multicolumn{1}{r}{0.616}\\ 
\multicolumn{1}{c}{$1.5$} & \multicolumn{1}{r}{0.056} & \multicolumn{1}{r}{0.057} & \multicolumn{1}{r}{0.048} & \multicolumn{1}{r}{0.054} & \multicolumn{1}{r}{1.000} & 
\multicolumn{1}{r}{1.000} & \multicolumn{1}{r}{0.878} & \multicolumn{1}{r}{1.000} & \multicolumn{1}{r}{1.000} & \multicolumn{1}{r}{0.960} & \multicolumn{1}{r}{0.635}\\  \hline\hline
\end{tabular}
\end{equation*}
\end{table}

\begin{table}[!htbp]
\caption{Empirical rejection rates of $CvM_n^{block}$ with $h=n^{-1/3.5}$ and $L=\lfloor an^{1/4} \rfloor$}\label{cvm_bmb} 
\begin{equation*}
\begin{tabular}{@{}clllllllllll}
\hline\hline
 & \multicolumn{11}{c}{DGPs} \\ \hline
\multicolumn{1}{c}{$a$} & (S1) & (S2) & (S3) & (S4) & (P1) & 
(P2) & (P3) & (P4) & (P5)& (P6)& (P7)\\ \hline
\multicolumn{12}{c}{} \\ 
\multicolumn{12}{c}{$n=100$} \\ 
&  &  &  &  &  &  & \\ 
\multicolumn{1}{c}{$1$} & \multicolumn{1}{r}{0.052} & \multicolumn{1}{r}{0.053} & \multicolumn{1}{r}{0.069} & \multicolumn{1}{r}{0.082} & \multicolumn{1}{r}{0.977} & 
\multicolumn{1}{r}{0.318} & \multicolumn{1}{r}{0.129} & \multicolumn{1}{r}{0.240} & \multicolumn{1}{r}{0.199} & \multicolumn{1}{r}{0.147} & \multicolumn{1}{r}{0.116}\\ 
\multicolumn{1}{c}{$2$} & \multicolumn{1}{r}{0.037} & \multicolumn{1}{r}{0.049} & \multicolumn{1}{r}{0.056} & \multicolumn{1}{r}{0.068} & \multicolumn{1}{r}{0.970} & 
\multicolumn{1}{r}{0.285} & \multicolumn{1}{r}{0.118} & \multicolumn{1}{r}{0.203} & \multicolumn{1}{r}{0.179} & \multicolumn{1}{r}{0.172} & \multicolumn{1}{r}{0.112}\\ 
\multicolumn{1}{c}{$4$} & \multicolumn{1}{r}{0.039} & \multicolumn{1}{r}{0.036} & \multicolumn{1}{r}{0.036} & \multicolumn{1}{r}{0.058} & \multicolumn{1}{r}{0.896} & 
\multicolumn{1}{r}{0.175} & \multicolumn{1}{r}{0.061} & \multicolumn{1}{r}{0.156} & \multicolumn{1}{r}{0.169} & \multicolumn{1}{r}{0.134} & \multicolumn{1}{r}{0.099}\\ 
&  &  &  &  &  &  & \\ 
\multicolumn{12}{c}{$n=200$} \\ 
&  &  &  &  &  &  & \\ 
\multicolumn{1}{c}{$1$} & \multicolumn{1}{r}{0.058} & \multicolumn{1}{r}{0.060} & \multicolumn{1}{r}{0.052} & \multicolumn{1}{r}{0.070} & \multicolumn{1}{r}{1.000} & 
\multicolumn{1}{r}{0.878} & \multicolumn{1}{r}{0.373} & \multicolumn{1}{r}{0.575} & \multicolumn{1}{r}{0.409} & \multicolumn{1}{r}{0.241} & \multicolumn{1}{r}{0.155}\\ 
\multicolumn{1}{c}{$2$} & \multicolumn{1}{r}{0.054} & \multicolumn{1}{r}{0.058} & \multicolumn{1}{r}{0.062} & \multicolumn{1}{r}{0.057} & \multicolumn{1}{r}{1.000} & 
\multicolumn{1}{r}{0.852} & \multicolumn{1}{r}{0.305} & \multicolumn{1}{r}{0.553} & \multicolumn{1}{r}{0.412} & \multicolumn{1}{r}{0.223} & \multicolumn{1}{r}{0.177}\\  
\multicolumn{1}{c}{$4$} & \multicolumn{1}{r}{0.037} & \multicolumn{1}{r}{0.054} & \multicolumn{1}{r}{0.052} & \multicolumn{1}{r}{0.068} & \multicolumn{1}{r}{1.000} & 
\multicolumn{1}{r}{0.838} & \multicolumn{1}{r}{0.254} & \multicolumn{1}{r}{0.515} & \multicolumn{1}{r}{0.396} & \multicolumn{1}{r}{0.249} & \multicolumn{1}{r}{0.115}\\ 
&  &  &  &  &  &  & \\ 
\multicolumn{12}{c}{$n=400$} \\ 
&  &  &  &  &  &  & \\ 
\multicolumn{1}{c}{$1$} & \multicolumn{1}{r}{0.059} & \multicolumn{1}{r}{0.071} & \multicolumn{1}{r}{0.049} & \multicolumn{1}{r}{0.064} & \multicolumn{1}{r}{1.000} & 
\multicolumn{1}{r}{1.000} & \multicolumn{1}{r}{0.769} & \multicolumn{1}{r}{0.973} & \multicolumn{1}{r}{0.937} & \multicolumn{1}{r}{0.641} & \multicolumn{1}{r}{0.312}\\ 
\multicolumn{1}{c}{$2$} & \multicolumn{1}{r}{0.053} & \multicolumn{1}{r}{0.061} & \multicolumn{1}{r}{0.067} & \multicolumn{1}{r}{0.073} & \multicolumn{1}{r}{1.000} & 
\multicolumn{1}{r}{0.998} & \multicolumn{1}{r}{0.726} & \multicolumn{1}{r}{0.976} & \multicolumn{1}{r}{0.955} & \multicolumn{1}{r}{0.662} & \multicolumn{1}{r}{0.319}\\ 
\multicolumn{1}{c}{$4$} & \multicolumn{1}{r}{0.047} & \multicolumn{1}{r}{0.042} & \multicolumn{1}{r}{0.062} & \multicolumn{1}{r}{0.062} & \multicolumn{1}{r}{1.000} & 
\multicolumn{1}{r}{0.999} & \multicolumn{1}{r}{0.706} & \multicolumn{1}{r}{0.959} & \multicolumn{1}{r}{0.896} & \multicolumn{1}{r}{0.613} & \multicolumn{1}{r}{0.301}\\ 
&  &  &  &  &  &  & \\ 
\multicolumn{12}{c}{$n=800$} \\ 
&  &  &  &  &  &  & \\ 
\multicolumn{1}{c}{$1$} & \multicolumn{1}{r}{0.046} & \multicolumn{1}{r}{0.056} & \multicolumn{1}{r}{0.047} & \multicolumn{1}{r}{0.058} & \multicolumn{1}{r}{1.000} & 
\multicolumn{1}{r}{1.000} & \multicolumn{1}{r}{0.993} & \multicolumn{1}{r}{1.000} & \multicolumn{1}{r}{1.000} & \multicolumn{1}{r}{0.998} & \multicolumn{1}{r}{0.828}\\ 
\multicolumn{1}{c}{$2$} & \multicolumn{1}{r}{0.040} & \multicolumn{1}{r}{0.048} & \multicolumn{1}{r}{0.057} & \multicolumn{1}{r}{0.057} & \multicolumn{1}{r}{1.000} & 
\multicolumn{1}{r}{1.000} & \multicolumn{1}{r}{0.992} & \multicolumn{1}{r}{1.000} & \multicolumn{1}{r}{1.000} & \multicolumn{1}{r}{0.998} & \multicolumn{1}{r}{0.829}\\ 
\multicolumn{1}{c}{$4$} & \multicolumn{1}{r}{0.046} & \multicolumn{1}{r}{0.056} & \multicolumn{1}{r}{0.058} & \multicolumn{1}{r}{0.066} & \multicolumn{1}{r}{1.000} & 
\multicolumn{1}{r}{1.000} & \multicolumn{1}{r}{0.991} & \multicolumn{1}{r}{1.000} & \multicolumn{1}{r}{1.000} & \multicolumn{1}{r}{0.992} & \multicolumn{1}{r}{0.813}\\  \hline\hline
\end{tabular}
\end{equation*}
\end{table}

\begin{table}[!htbp]
\caption{Empirical rejection rates of $KS_n^{block}$ with $h=n^{-1/3.5}$ and $L=\lfloor an^{1/4} \rfloor$}\label{ks_bmb} 
\begin{equation*}
\begin{tabular}{@{}clllllllllll}
\hline\hline
 & \multicolumn{11}{c}{DGPs} \\ \hline
\multicolumn{1}{c}{$a$} & (S1) & (S2) & (S3) & (S4) & (P1) & 
(P2) & (P3) & (P4) & (P5)& (P6)& (P7)\\ \hline
\multicolumn{12}{c}{} \\ 
\multicolumn{12}{c}{$n=100$} \\ 
&  &  &  &  &  &  & \\ 
\multicolumn{1}{c}{$1$} & \multicolumn{1}{r}{0.044} & \multicolumn{1}{r}{0.032} & \multicolumn{1}{r}{0.040} & \multicolumn{1}{r}{0.060} & \multicolumn{1}{r}{0.814} & 
\multicolumn{1}{r}{0.274} & \multicolumn{1}{r}{0.096} & \multicolumn{1}{r}{0.224} & \multicolumn{1}{r}{0.200} & \multicolumn{1}{r}{0.139} & \multicolumn{1}{r}{0.082}\\ 
\multicolumn{1}{c}{$2$} & \multicolumn{1}{r}{0.037} & \multicolumn{1}{r}{0.035} & \multicolumn{1}{r}{0.032} & \multicolumn{1}{r}{0.054} & \multicolumn{1}{r}{0.760} & 
\multicolumn{1}{r}{0.227} & \multicolumn{1}{r}{0.093} & \multicolumn{1}{r}{0.207} & \multicolumn{1}{r}{0.163} & \multicolumn{1}{r}{0.122} & \multicolumn{1}{r}{0.074}\\ 
\multicolumn{1}{c}{$4$} & \multicolumn{1}{r}{0.029} & \multicolumn{1}{r}{0.022} & \multicolumn{1}{r}{0.028} & \multicolumn{1}{r}{0.041} & \multicolumn{1}{r}{0.546} & 
\multicolumn{1}{r}{0.130} & \multicolumn{1}{r}{0.056} & \multicolumn{1}{r}{0.140} & \multicolumn{1}{r}{0.133} & \multicolumn{1}{r}{0.094} & \multicolumn{1}{r}{0.055}\\ 
&  &  &  &  &  &  & \\ 
\multicolumn{12}{c}{$n=200$} \\ 
&  &  &  &  &  &  & \\ 
\multicolumn{1}{c}{$1$} & \multicolumn{1}{r}{0.043} & \multicolumn{1}{r}{0.044} & \multicolumn{1}{r}{0.042} & \multicolumn{1}{r}{0.053} & \multicolumn{1}{r}{0.996} & 
\multicolumn{1}{r}{0.661} & \multicolumn{1}{r}{0.206} & \multicolumn{1}{r}{0.543} & \multicolumn{1}{r}{0.391} & \multicolumn{1}{r}{0.211} & \multicolumn{1}{r}{0.136}\\ 
\multicolumn{1}{c}{$2$} & \multicolumn{1}{r}{0.048} & \multicolumn{1}{r}{0.043} & \multicolumn{1}{r}{0.039} & \multicolumn{1}{r}{0.050} & \multicolumn{1}{r}{0.988} & 
\multicolumn{1}{r}{0.610} & \multicolumn{1}{r}{0.182} & \multicolumn{1}{r}{0.500} & \multicolumn{1}{r}{0.344} & \multicolumn{1}{r}{0.204} & \multicolumn{1}{r}{0.139}\\  
\multicolumn{1}{c}{$4$} & \multicolumn{1}{r}{0.038} & \multicolumn{1}{r}{0.043} & \multicolumn{1}{r}{0.041} & \multicolumn{1}{r}{0.042} & \multicolumn{1}{r}{0.969} & 
\multicolumn{1}{r}{0.532} & \multicolumn{1}{r}{0.135} & \multicolumn{1}{r}{0.429} & \multicolumn{1}{r}{0.309} & \multicolumn{1}{r}{0.194} & \multicolumn{1}{r}{0.104}\\ 
&  &  &  &  &  &  & \\ 
\multicolumn{12}{c}{$n=400$} \\ 
&  &  &  &  &  &  & \\ 
\multicolumn{1}{c}{$1$} & \multicolumn{1}{r}{0.046} & \multicolumn{1}{r}{0.050} & \multicolumn{1}{r}{0.042} & \multicolumn{1}{r}{0.055} & \multicolumn{1}{r}{1.000} & 
\multicolumn{1}{r}{0.987} & \multicolumn{1}{r}{0.416} & \multicolumn{1}{r}{0.940} & \multicolumn{1}{r}{0.782} & \multicolumn{1}{r}{0.484} & \multicolumn{1}{r}{0.269}\\ 
\multicolumn{1}{c}{$2$} & \multicolumn{1}{r}{0.051} & \multicolumn{1}{r}{0.042} & \multicolumn{1}{r}{0.058} & \multicolumn{1}{r}{0.063} & \multicolumn{1}{r}{1.000} & 
\multicolumn{1}{r}{0.980} & \multicolumn{1}{r}{0.400} & \multicolumn{1}{r}{0.940} & \multicolumn{1}{r}{0.770} & \multicolumn{1}{r}{0.496} & \multicolumn{1}{r}{0.240}\\ 
\multicolumn{1}{c}{$4$} & \multicolumn{1}{r}{0.034} & \multicolumn{1}{r}{0.039} & \multicolumn{1}{r}{0.051} & \multicolumn{1}{r}{0.039} & \multicolumn{1}{r}{1.000} & 
\multicolumn{1}{r}{0.970} & \multicolumn{1}{r}{0.375} & \multicolumn{1}{r}{0.888} & \multicolumn{1}{r}{0.727} & \multicolumn{1}{r}{0.409} & \multicolumn{1}{r}{0.228}\\ 
&  &  &  &  &  &  & \\ 
\multicolumn{12}{c}{$n=800$} \\ 
&  &  &  &  &  &  & \\ 
\multicolumn{1}{c}{$1$} & \multicolumn{1}{r}{0.054} & \multicolumn{1}{r}{0.044} & \multicolumn{1}{r}{0.039} & \multicolumn{1}{r}{0.050} & \multicolumn{1}{r}{1.000} & 
\multicolumn{1}{r}{1.000} & \multicolumn{1}{r}{0.859} & \multicolumn{1}{r}{1.000} & \multicolumn{1}{r}{0.999} & \multicolumn{1}{r}{0.925} & \multicolumn{1}{r}{0.571}\\ 
\multicolumn{1}{c}{$2$} & \multicolumn{1}{r}{0.050} & \multicolumn{1}{r}{0.025} & \multicolumn{1}{r}{0.054} & \multicolumn{1}{r}{0.050} & \multicolumn{1}{r}{1.000} & 
\multicolumn{1}{r}{1.000} & \multicolumn{1}{r}{0.822} & \multicolumn{1}{r}{1.000} & \multicolumn{1}{r}{0.999} & \multicolumn{1}{r}{0.907} & \multicolumn{1}{r}{0.582}\\ 
\multicolumn{1}{c}{$4$} & \multicolumn{1}{r}{0.042} & \multicolumn{1}{r}{0.046} & \multicolumn{1}{r}{0.050} & \multicolumn{1}{r}{0.055} & \multicolumn{1}{r}{1.000} & 
\multicolumn{1}{r}{1.000} & \multicolumn{1}{r}{0.822} & \multicolumn{1}{r}{1.000} & \multicolumn{1}{r}{0.998} & \multicolumn{1}{r}{0.895} & \multicolumn{1}{r}{0.554}\\  \hline\hline
\end{tabular}
\end{equation*}
\end{table}

\begin{table}[!htbp]
\caption{Testing for nonlinear predicability from VRP to RP}\label{empirical} 
\small
\begin{equation*}
\begin{tabular}{lcccccc}
\hline\hline
Direction of Causality & $h=cn^{-1/3.5}$ & 
$CvM_n$ & $KS_n$ & $CvM^{block}_n$ & $KS^{block}_n$ & LIN\\ 
\hline
\multicolumn{7}{c}{\textbf{Horizon: One Month}} \\ 
&  &  &  &\\ 
& \multicolumn{1}{l}{$c=0.5$} & 0.2267 & 0.3377 & 0.2298 & 0.3054 \\ 
\textit{VRP}$\rightarrow $\textit{RP} & \multicolumn{1}{l}{$c=1.0$} & 0.1711 & 0.1343 & 0.1764 & 0.1300 & 0.0017 (0.4172)\\ 
& \multicolumn{1}{l}{$c=1.5$} & 0.1077 & 0.0585 & 0.1073 & 0.0564\\ 
&  &  &  &\\ 
\multicolumn{7}{c}{\textbf{Horizon: Three Months}} \\ 
&  &  &  &\\ 
& \multicolumn{1}{l}{$c=0.5$} & 0.0640 & 0.0663 & 0.0642 & 0.0682 \\ 
\textit{VRP}$\rightarrow $\textit{RP} & \multicolumn{1}{l}{$c=1.0$} & 0.0141 & 0.0109 & 0.0202 & 0.0201 & 0.0016 (1.5009)\\ 
& \multicolumn{1}{l}{$c=1.5$} & 0.0056 & 0.0055 & 0.0090 & 0.0142\\ 
&  &  &  &\\ 
\multicolumn{7}{c}{\textbf{Horizon: Six Months}} \\ 
&  &  &  &\\ 
& \multicolumn{1}{l}{$c=0.5$} & 0.0000 & 0.0001 & 0.0011 & 0.0022 \\ 
\textit{VRP}$\rightarrow $\textit{RP} & \multicolumn{1}{l}{$c=1.0$} & 0.0000 & 0.0000 & 0.0002 & 0.0013 & 0.0002 (0.3329)\\ 
& \multicolumn{1}{l}{$c=1.5$} & 0.0000 & 0.0000 & 0.0000 & 0.0027\\ 
&  &  &  &\\ 
\multicolumn{7}{c}{\textbf{Horizon: Nine Months}} \\ 
& & & &\\
& \multicolumn{1}{l}{$c=0.5$} & 0.0000 & 0.0000 & 0.0001 & 0.0004 \\ 
\textit{VRP}$\rightarrow $\textit{RP} & \multicolumn{1}{l}{$c=1.0$} & 0.0000 & 0.0000 & 0.0000 & 0.0004 & 0.0005 (2.0807)\\ 
& \multicolumn{1}{l}{$c=1.5$} & 0.0000 & 0.0000 & 0.0000 & 0.0004\\   \hline\hline
\end{tabular}%
\end{equation*}
\noindent \textbf{Note}$\text{:}$ {\small For tests $CvM_n$, $KS_n$, $CvM_n^{block}$ and $KS_n^{block}$, (block) multiplier bootstrap $p$-values are reported. For the block tests, $L=\lfloor 2n^{1/4} \rfloor$ is used. LIN corresponds to the linear test, where the least squares estimate $\hat{\alpha}_{\tau}$ and its corresponding $t$-statistic $t_{\hat{\alpha}_{\tau}}$ (in parentheses) based on HAC robust variance estimator, are reported.}

\end{table}

\end{document}